\documentclass[aoas ,preprint]{imsart}

\RequirePackage[OT1]{fontenc}
\RequirePackage{amsthm,amsmath}
\RequirePackage{natbib}
\RequirePackage[colorlinks,citecolor=blue,urlcolor=blue]{hyperref}
\usepackage{subfigure}
\usepackage{float}
\usepackage{graphicx}
\usepackage{tikz}
\usetikzlibrary{fit,positioning,calc}
\usepackage[margin=1.2in,footskip=0.25in]{geometry}
\usepackage{amssymb}
\def\T{{ \mathrm{\scriptscriptstyle T} }}

\arxiv{arXiv:1505.05668v2}
\bibpunct{[}{]}{,}{()}{}{}
\startlocaldefs
\numberwithin{equation}{section}
\theoremstyle{plain}

\endlocaldefs
\begin{document}

\begin{frontmatter}
\title{Locally Adaptive Dynamic Networks}
\runtitle{Locally Adaptive Dynamic Networks}

\begin{aug}
\author{\fnms{Daniele} \snm{Durante}\thanksref{m1}\ead[label=e1]{durante@stat.unipd.it}}
\and
\author{\fnms{David B.} \snm{Dunson}\thanksref{m2}
\ead[label=e3]{dunson@duke.edu}
}

\runauthor{Durante and Dunson}

\affiliation{University of Padova\thanksmark{m1} and Duke University\thanksmark{m2}}

\address{University of Padova, \\ Department of Statistical Sciences.\\
Via Cesare Battisti, 241, 35121 Padua, Italy.\\
\printead{e1}}

\address{Duke University,\\
Department of Statistical Science.\\
Box 90251, Durham, NC 27708 34127, USA.\\
\printead{e3}\\
}
\end{aug}

\begin{abstract}
Our focus is on realistically modeling and forecasting dynamic networks of face-to-face contacts among individuals.  Important aspects of such data that lead to problems with current methods include the tendency of the contacts to move between periods of slow and rapid changes, and the dynamic  heterogeneity in the actors' connectivity behaviors.   Motivated by this application, we develop a novel method for Locally Adaptive DYnamic (LADY) network inference.  The proposed model relies on a dynamic latent space representation in which each actor's position evolves in time via stochastic differential equations.  Using a state space representation for these stochastic processes and P\'olya-gamma data augmentation, we develop an efficient MCMC algorithm for posterior inference along with tractable procedures for online updating and forecasting of future networks.  We evaluate performance in simulation studies, and consider an application to face-to-face contacts among individuals in a primary school.
\end{abstract}

\begin{keyword}
\kwd{Face-to-Face Dynamic Contact Network}
\kwd{Latent Space}
\kwd{Nested Gaussian Process}
\kwd{Online Updating}
\kwd{P\`olya-Gamma}
\kwd{State Space Model}
\end{keyword}

\end{frontmatter}

\section{Introduction} \label{intro}
\begin{figure}
\centering
\includegraphics[trim=0.8cm 0.85cm 0cm 0.1cm, clip=true,height=5cm, width=16cm]{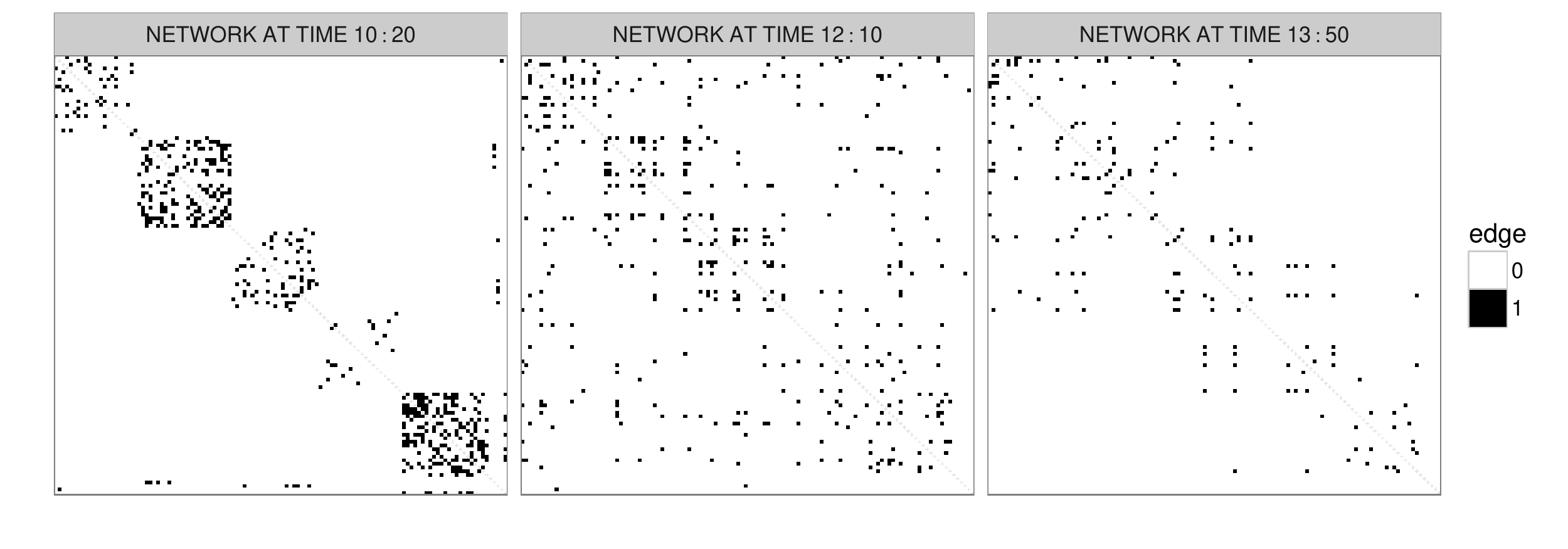}
\caption{\footnotesize{For selected times, adjacency matrices representing the observed face-to-face contact networks.}}
\label{Motivating}
\end{figure}

We are interested in studying face-to-face dynamic interactions among individuals in a primary school. Understanding key aspects of these time-varying interaction networks and forecasting of future contacts is interesting sociologically and important in infectious disease epidemiology.  As illustrated in Figure \ref{Motivating}, data consist of a sequence of $V \times V$ time-varying symmetric adjacency matrices $Y_{t_1}, \ldots, Y_{t_n}$ having entries $Y_{t_i[vu]}=Y_{t_i[uv]}=1$ if a face-to-face contact has been recorded between actors $v=2, \ldots, V$ and $u=1, \ldots, v-1$ at time $t_i=t_1, \ldots, t_n$, and $Y_{t_i[vu]}=Y_{t_i[uv]}=0$ if no contact has been observed. These undirected dynamic networks are available at \url{http://www.sociopatterns.org}; see also  \citet{steh_2011} and \citet{gem_2014} for additional details. 

The increasing availability of new sensing devices and wearable sensors to trace human interaction behaviors allows a growing access to these type of dynamic networks, while opening new avenues for studying underlying patterns in social interactions and how these processes relate to associated dynamic systems such as epidemic spreading. Recent studies have investigated dynamic face-to-face human interactions in several environments. \citet{ise_2011} focus on contact dynamics among individuals in two different scenarios, covering a scientific conference and a long-running museum exhibition, respectively. \citet{van_2013} study interactions among staff members and patients in a hospital.  \citet{steh_2011}, \citet{gem_2014} and \citet{four_2014}, \citet{mas_2015} investigate face-to-face contact dynamics among students in primary and high schools, respectively; refer also to \citet{bar_2013} for a review.

The above studies mostly focus on descriptive analyses in order to provide a summarized overview of the topological structures  underlying the observed networks and how these measures relate to  environmental conditions and other variables. \citet{wyat_2008} analyze, instead, face-to-face contacts  from a model-based perspective, aggregating the dynamic interactions into a single network  measuring duration of contact. Although these procedures provide valuable insights, flexible statistical models of how the human interaction networks evolve in time would provide improved ability to jointly infer dynamic changes in network structures, while accounting for uncertainty.  In addition, such models would be useful in terms of prediction and forecasting of contacts, which is of key interest in epidemiology.  

There is a rich literature on dynamic networks, but several aspects of our motivating application require careful innovation.  These aspects include the tendency of the contacts to move between periods of slow and rapid changes, the dynamic heterogeneity in the actors' connectivity behaviors, and the need for fast and accurate online updating and forecasting procedures for timely prediction of future interactions and design of appropriate outbreak prevention policies. In addressing these goals, it is fundamental to borrow information within each network and across time, while scaling to a larger number of time points $t_1, \ldots, t_n$ and to a moderately large set of actors, without affecting flexibility in modeling dynamic contacts.

\subsection{Dynamic face-to-face human contact networks} \label{data}
We focus on the face-to-face dynamic interaction networks among individuals in a primary school in Lyon, France. Raw contact data are available  for $232$ students between 6 and 12 years of age and $10$ teachers, during two consecutive school days running from $\sim08{:}40$ to $\sim17{:}10$. The primary school is characterized by $5$ grades, each divided into two classes comprising on average $24$ students.

Face-to-face contact data are monitored via wearable radio frequency identification devices (RFID), exchanging low-power radio packets when two individuals face each other at a distance of $\sim1-1.5$ meters. This proximity  range is chosen to represent a reasonable proxy of close social contact, while indicating a potential occasion of disease transmission \citep{steh_2011}. Raw data are available for consecutive windows of $20$ seconds and encode which pairs of individuals had a face-to-face  contact in each one of these time intervals; refer to \citet{cat_2010} for a description of RFID proximity-sensing devices.

Initial descriptive analyses of these data highlight a very sparse and noisy structure with only $40$ contacts --- among the $29{,}161$ possible --- monitored on average for every window of $20$ seconds. This time scale might be too narrow to highlight recurring patterns in the dynamic evolution of the underlying network topological structures. Hence, we aggregate the data in consecutive time windows of $10$ minutes so that the resulting networks encode which pairs of individuals established at least one face-to-face proximity contact during each one of these consecutive $10$ minutes time intervals. Focusing on binary connections instead of the cumulative number of contacts in the $10$ minutes time windows provides a simpler starting point. Moreover, under an  epidemiological perspective, at least one proximity contact of $20$ seconds may be sufficient for disease transmission.  Although we loose  short scale dynamics, these windows are sufficiently wide to highlight longer range patterns in the network topology, but maintain enough granularity to capture sharp changes which may occur in correspondence of breaks, lunch times and school hours. We found these underlying structures quite robust to moderate changes in the length of the time intervals, including $5$, $15$ and $20$ minutes. \citet{steh_2011} consider a similar aggregation to study dynamic changes in the averaged degree.

\begin{figure}
\centering
\includegraphics[trim=0.6cm 0.5cm 0.2cm 0.1cm, clip=true,height=6.6cm, width=15.5cm]{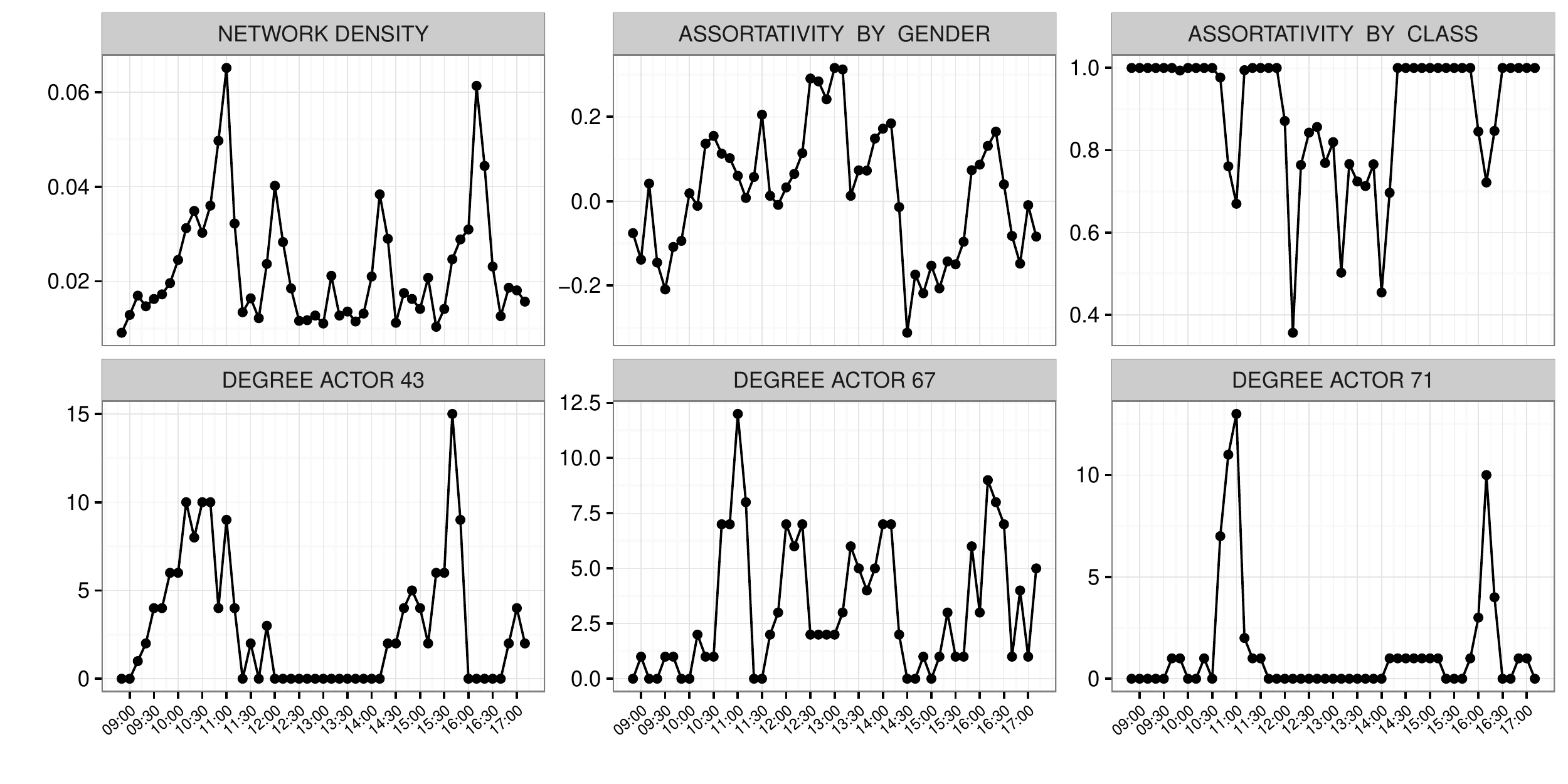}
\caption{\footnotesize{Time-varying observed network summary measures for the first day of school. Upper panels: global network measures. Bottom panels: degree of selected actors.}}
\label{Motivating1}
\end{figure}

In analyzing these data, we seek inference and forecasting procedures which are sufficiently flexible to capture different types of dynamic changes in the network data.  Dynamic variations in connectivity patterns may be influenced by the underlying endogenous architectures as well as exogenous factors, such as changing spatial environments and class or gender homophily. Information on class membership and gender are available for all the individuals --- except teachers --- while approximate changes in spatio-temporal locations are provided for $5$ classes out of $10$ in Figure 10 of  \citet{steh_2011}.  We focus on the students and teachers in these 5 classes --- for a total of $V=120$ actors --- and provide inference on the data from the first day  $Y_{t_1}, \ldots, Y_{t_n}$, while using networks  $Y^*_{t_1}, \ldots, Y^*_{t_n}$ in the second day to evaluate out-of-sample predictive performance.  In analyzing these data, we develop a flexible dynamic latent space model relaxing the complex dependence structure within the network to one of conditional independence between contacts among pairs of actors given their positions in a latent space. Therefore, focusing on a subset of individuals of interest, instead of modeling the network of contacts in the entire school, does not contradict the assumptions of our statistical model.

As shown in Figure \ref{Motivating1}, the trajectories of the global and actor-specific summary measures cycle irregularly between phases characterized by slower and more rapid variations. Flexibly capturing such behavior is important  to improve prediction and investigate how dynamic face-to-face interactions relate to specific events, such as school hours, breaks, lunch time and changing environments. Instead of directly including covariate information on gender, class membership and spatial locations in the dynamic model, we use these information to assess the extent to which our model can learn known structures in the data.  Current models for dynamic networks typically rely on homogeneity and stationarity assumptions --- for both endogenous and exogenous effects --- and hence have difficulties in modeling variations over time of specific network structures.  This can have a strong effect on the quality of inferences and predictions, with under-smoothing during periods of stable contacts and over-smoothing across times of rapid variations. Motivated by our face-to-face contact network data and by the need for flexible methods enforcing time-varying smoothness in dynamic networks, we develop a Locally Adaptive DYnamic (LADY) network model that characterizes the time-varying edge probabilities via latent processes, which have time-varying smoothness.

\subsection{Relevant literature} \label{lite}
There is a growing literature on statistical modeling of dynamic networks.  Much of the literature focuses on the case in which the exact time of each contact event is observed;  see for example \citet{butts_2008} and \citet{dub_2013}.  We instead consider the case in which snapshots of a dynamic network are observed at different time points.  

A popular class of procedures generalizes exponential random graph models (ERGMs) to include discrete time Markov dynamics \citep{rob_2001, hann_2007, kriv_2013}.   \citet{holl_1977}, and the subsequent improvements of  \citet{snij_2001, snij_2005} and \citet{snij_2010} provide alternative continuous time Markov specifications allowing the rate of change in the network to vary with time.  These models are elegant and allow dynamic inference on several network characteristics.  However, current specifications do not  fully accommodate non-homogeneous dynamics and heterogenous contact patterns, which are key aspects of our data.

There is also a rich literature on alternatives to ERGMs, including stochastic block models \citep{fien_1981,now_2001},  mixed membership stochastic block models \citep{air_2008} and latent space models \citep{hof_2002}.  These methods characterize the edges as conditionally independent Bernoulli variables given their corresponding edge probabilities, with these probabilities further defined as a function of actor-specific latent variables. As highlighted in \citet{hun_2012}, building on conditional independence provides computational benefits in facilitating implementation of MCMC methods. Moreover, although --- differently from dynamic ERGMs --- these procedures do not explicitly parameterize inter-dependence between relations, the shared dependence on a common set of actor-specific latent variables can induce flexible dependence structures and allow for across-actor heterogeneity in dynamic contacts.

Dynamic stochastic block models have been considered by \citet{yang_2009, yang_2010} and later refined by  \citet{xu_2014} and \citet{xu_2015}. These approaches are specifically tailored for learning dynamic changes in shared connectivity behaviors, and may fail to accurately characterize and predict contact patterns different than those arising from block structures. Dynamic relational feature models   \citep{foul_2011} improve flexibility by replacing the single block membership indicator with vectors describing presence or absence of features for each actor, but assume a time-constant features-interaction matrix restricting the dynamics.

Dynamic mixed membership stochastic block models \citep{xing_2010} and latent space models \citep{sark_2005, sew_2015} are more flexible.  Typical approaches incorporate dynamics through state space models, Markov processes and random walk trajectories.  To address computational intractability, approximations are used including the extended Kalman filter and variational approaches.  \citet{dur_2014} embed the actors in a latent  space and allow their coordinates to evolve in continuous time via Gaussian processes (GP).  Their approach  provides a simple Gibbs sampling algorithm, but faces the usual computational bottlenecks of GPs in scaling to a large numbers of time points, and the dynamic network inherits the stationary dependence structure of the latent GPs.

Outside of the network field, there are several approaches to incorporate locally adaptive smoothness in dynamic processes.  Particularly relevant to our work is the nested Gaussian process (nGP) proposed by \citet{zhu_2013} for regression and extended to multivariate time series by \citet{dur_JMLR_2014}.  The nGP models the trajectories' $m$th order derivatives via GPs, which are in turn centered on a higher level GP instantaneous mean that favors time-varying smoothness.  Similar constructions are lacking in network fields.
 
Our LADY network model accounts for across-actor heterogeneity via a latent space formulation with nGPs to induce time variations in the rate of change of the network structure.  By considering a state space representation for the latent processes, we reduce the computational burden from cubic in the number of time points to linear, while  developing simple procedures for forecasting, prediction and online updating appropriate to streaming networks. In Section \ref{mod} we describe the LADY network model. Posterior computation procedures are provided in Section \ref{post} with an additional focus on forecasting and online updating. In Section \ref{sim} we consider a simulation study to test our methods in relation to existing dynamic network models. Section \ref{app} presents the results for our analysis of face-to-face contact network data.

\section{LADY networks} \label{mod}
We assume that the observed data $Y_{t_1}, \ldots, Y_{t_n}$ provide a realization --- on a discrete time grid $t_1, \ldots, t_n$ --- of the continuous latent process $\{\mathcal{Y}(t): t \in \mathbb{T} \subset \Re^+ \}$ and seek a representation for the generative mechanism associated with this network-valued stochastic process, which is consistent with the goals discussed in Section \ref{intro}. As our contact networks are undirected, it is sufficient to model the lower triangular elements of $\mathcal{Y}(t)$ since $\mathcal{Y}_{vu}(t)=\mathcal{Y}_{uv}(t)$ for every $v=2, \ldots, V$, $u=1, \ldots, v-1$ and $t\in \mathbb{T}$. In particular, we let
\begin{eqnarray}
\mathcal{Y}_{vu}(t) \mid \pi_{vu}(t) \sim \mbox{Bern}\left\{\pi_{vu}(t)\right\},
\label{eq1}
\end{eqnarray}
for every $v=2,\ldots,V$, $u=1,\ldots,v-1$ and $t \in \mathbb{T}$, with  
\begin{eqnarray}
\pi_{vu}(t)=\big[1+\exp\{-\mu(t)-x_{v}(t)^{\T}x_{u}(t)\}\big]^{-1}.
\label{eq1.1}
\end{eqnarray}
Under \eqref{eq1} the edges $\mathcal{Y}_{vu}(t) \in \{0, 1 \}$ are conditionally independent Bernoulli random variables given their corresponding edge probabilities $\pi_{vu}(t) \in (0,1)$.  These edge probabilities are obtained by  
mapping a latent similarity measure $s_{vu}(t)=\mu(t)+x_{v}(t)^{\T}x_{u}(t)$ from $\Re$ to $(0,1)$ according to \eqref{eq1.1}, where $\mu(t)$ represents a baseline similarity score shared by all the edges at time $t$, whereas $x_v(t) = \{x_{v1}(t),\ldots,x_{vH}(t)\}^{\T} \in \Re^H$ and  $x_u(t) = \{x_{u1}(t),\ldots,x_{uH}(t)\}^{\T} \in \Re^H$ are the vectors of latent coordinates for actors $v$ and $u$, respectively, at time $t$.  Based on \eqref{eq1.1}, the probability of an edge between actors $v$ and $u$ at time $t$ increases with $x_{v}(t)^{\T}x_{u}(t) \in \Re$.  

Our dynamic latent space representation collapses higher-order dependencies into a lower-dimensional space, reducing dimensionality from $V(V-1)/2$ stochastic processes on the edge probabilities to $V \times H$  --- typically $H \ll V$ --- latent trajectories and one baseline process.  This construction recalls the statistic eigenmodel in \citet{hoff_2007}, which provides a flexible class of latent variable constructions for social networks allowing across-actor heterogeneity and accommodating various topological properties.  Following \citet{hoff_2007}, model \eqref{eq1}--\eqref{eq1.1} generalizes stochastic block models  \citep{fien_1981,now_2001} and latent distance models \citep{hof_2002}, and can accommodate block structures, homophily, and transitive contact patterns. These properties are --- potentially --- key factors underlying our face-to-face interaction data. For instance, during school hours or lunch times, the contact networks may exhibit block structures due to shared environments by students belonging to the same class or group of classes. Breaks are instead potentially associated with transitive patterns arising from friendship among students in different classes or homophily by gender. The low-rank decomposition in \eqref{eq1.1} is not unique. However, we avoid identifiability restrictions on the latent coordinates, as they are not required to ensure the identifiability of the edge probabilities, which are the focus of prediction and inference. 

In order to complete a representation of the LADY network model, we require priors on the stochastic processes $\{\mu(t): t \in \mathbb{T}\}$ and  $\{x_{vh}(t): t \in \mathbb{T}\}$ for each $v=1,\ldots,V$ and $ h=1,\ldots,H$.  If we define stationary processes, which assume that the correlation between the realizations at times $t_i$ and $t_j$ only depends on the time lag $|t_i-t_j|$, it is straightforward to show that the resulting network-valued stochastic process will inherit this stationarity. Recalling the discussion in Section \ref{intro} and the descriptive analyses in Figure \ref{Motivating1}, it is necessary to accommodate non-stationarity to realistically characterize the dynamics underlying the face-to-face interaction data.  However, this needs to be done in a careful way to avoid needing to estimate many parameters related to non-stationarity and face computational intractability.  Although there is a rich literature on incorporating non-stationarity in GPs, such models tend to be highly challenging to implement even in simpler settings.  

With these issues in mind, we rely on nested GPs  \citep{zhu_2013} --- rather than GPs --- to induce highly flexible stochastic processes on $\{ \mu(t): t \in \mathbb{T}\}$ and $\{ x_{vh}(t): t \in \mathbb{T}\}$ for each $v=1,\ldots,V$ and $h=1,\ldots,H$.  The nGPs explicitly incorporate time-varying smoothness by defining stochastic differential equations for the function's derivatives. Focusing on the trajectory $\{x_{vh}(t): t \in \mathbb{T}\}$, the stochastic differential equation representation for the nGP can be accurately characterized by the following state equations for  $\{x_{vh}(t): t \in \mathbb{T}\}$, it's first order derivative $\{x_{vh}'(t): t \in \mathbb{T}\}$ and the  instantaneous mean $\{m_{vh}(t): t \in \mathbb{T}\}$
\renewcommand{\arraystretch}{0.8}
\begin{eqnarray}
\left[ \begin{array}{c}
x_{vh}(t_{i+1})\\
x'_{vh}(t_{i+1})\\
m_{vh}(t_{i+1}) \end{array} \right]&=&
\left[ \begin{array}{ccc}
1&\delta_{i}&0\\
0 &1&\delta_{i}\\
0 &0&1 \end{array} \right]
\left[ \begin{array}{c}
x_{vh}(t_{i})\\
x_{vh}'(t_{i})\\
m_{vh}(t_{i}) \end{array} \right]+
\left[ \begin{array}{cc}
0&0\\
1&0\\
0&1 \end{array} \right]
\left[ \begin{array}{c}
\eta_{x_{vh},i}\\
\eta_{m_{vh},i}\\
\end{array} \right],\label{eq:2}\\
&=& T_i
\left[ \begin{array}{c}
x_{vh}(t_{i})\\
x_{vh}'(t_{i})\\
m_{vh}(t_{i}) \end{array} \right]+
R_i
\left[ \begin{array}{c}
\eta_{x_{vh},i}\\
\eta_{m_{vh},i}\\
\end{array} \right] \nonumber, \quad i=1, \ldots, n,
\end{eqnarray}
\renewcommand{\arraystretch}{0.5}independently for $v=1, \ldots, V$ and $h=1, \ldots, H$, with $(\eta_{x_{vh},i}, \eta_{m_{vh},i})^{\T}\sim \mbox{N}_{2}(0,\Sigma_{vh,i})$, $\Sigma_{vh,i}=\mbox{diag}(\sigma^2_{x_{vh}} \delta_{i}, \sigma^2_{m_{vh}} \delta_{i})$ and $\delta_{i}=t_{i+1}-t_{i}$ sufficiently small. Similarly for $\{\mu(t): t \in \mathbb{T}\}$, we let
\renewcommand{\arraystretch}{0.8}
\begin{eqnarray}
\left[ \begin{array}{c}
\mu(t_{i+1})\\
\mu'(t_{i+1})\\
z(t_{i+1}) \end{array} \right]&=&
\left[ \begin{array}{ccc}
1&\delta_{i}&0\\
0 &1&\delta_{i}\\
0 &0&1 \end{array} \right]
\left[ \begin{array}{c}
\mu(t_{i})\\
\mu'(t_{i})\\
z(t_{i}) \end{array} \right]+
\left[ \begin{array}{cc}
0&0\\
1&0\\
0&1 \end{array} \right]
\left[ \begin{array}{c}
\eta_{\mu,i}\\
\eta_{z,i}\\
\end{array} \right], \label{eq:3}\\
&=& T_i
\left[ \begin{array}{c}
\mu(t_{i})\\
\mu'(t_{i})\\
z(t_{i}) \end{array} \right]+
R_i
\left[ \begin{array}{c}
\eta_{\mu,i}\\
\eta_{z,i}\\
\end{array} \right] \nonumber, \quad i=1, \ldots, n,
\end{eqnarray}
\renewcommand{\arraystretch}{0.5} where $(\eta_{\mu,i}, \eta_{z,i})^{\T}\sim \mbox{N}_{2}(0,\Sigma_{\mu,i})$, with $\Sigma_{\mu,i}=\mbox{diag}(\sigma^2_{\mu} \delta_{i}, \sigma^2_{z} \delta_{i})$.

The state equations \eqref{eq:2}--\eqref{eq:3} along with the observation equations  \eqref{eq1}--\eqref{eq1.1} induce a provably flexible nonlinear logistic state space model for adaptive dynamic network inference, which is defined at every time grid $t_1, \ldots, t_n$, including unequally spaced ones. Although the above state equations can be easily extended to model higher order derivatives for the latent trajectories and their local instantaneous means, equations \eqref{eq:2}--\eqref{eq:3} prove to be sufficiently flexible in inducing adaptive dynamics according to our results. 

There exists other possible methods for accommodating local adaptivity in the latent trajectories, such as free knot splines \citep{fried_1991}.  However, such approaches are computationally intensive due to the unknown numbers and locations of knots \citep{fried_1991, geo_1993}.  This creates particular problems for large time grids and applications requiring lots of changes.  Our approach is appealing in providing a simple state space representation, which characterizes the latent positions at time $t_{i+1}$ as a  first-order stochastic Taylor expansion of the same quantities at $t_{i}$. This choice improves scalability of the inference procedures, while facilitating implementation of fast and tractable online updating  and prediction strategies by adapting available techniques for state space models.

\section{Bayesian inference} \label{post} Let $\Pi_{\pi}$ be the prior for $\{\pi_{vu}(t): v=2, \ldots, V, \ u=1, \ldots, v-1, \ t \in \mathbb{T} \}$ induced via  \eqref{eq1.1} by the state equations  \eqref{eq:2}--\eqref{eq:3} characterizing the nGP  priors $\Pi_{\mu}$ and $\Pi_{X}$, respectively. We consider a Bayesian approach for inference to update $\Pi_{\pi}$ given  the observed data $Y_{t_1}, \ldots, Y_{t_n}$. We leverage the P\`olya-gamma data augmentation for Bayesian logistic regression; see  \citet{pol_2013} for  details and \citet{choi_2013} for theoretical properties. Letting $\mathcal{Y}_i \sim \mbox{Bern}(\pi_i)$ independently, with $\pi_i = (1+e^{-x_i^{\T}\beta})^{-1}$,  \citet{pol_2013} show that conditionally on P\`olya-gamma augmented data $\omega_i \sim \mbox{PG}(1, x_i^{\T} \beta)$, the contribution to the likelihood  for the $i$th observation $y_i \in \{0,1 \}$ is
\begin{eqnarray}
\propto \exp \left[-\frac{\omega_i}{2}\left\{(y_i-0.5)/\omega_i-x_i^{\T}\beta \right\}^2 \right], \quad i=1, \ldots, n.
\label{eq5}
\end{eqnarray}
Equation \eqref{eq5} is the kernel of a Gaussian distribution for data $(y_i-0.5)/\omega_i$, with mean $x_i^{\T}\beta$ and variance $1/\omega_i$. As a result, letting $\beta \sim \mbox{N}_p(b,B)$ be the prior for the coefficient vector $\beta$, given P\`olya-gamma augmented data, the Bayesian logistic regression on data $y_i$ can be recast in terms of  Bayesian linear regression with Gaussian transformed response $(y_i-0.5)/\omega_i$. This allows a Gibbs algorithm, which alternates between sampling P\`olya-gamma augmented data and updating the coefficient vector $\beta$ from its full conditional Gaussian distribution.

By exploiting these results, we develop a simple and efficient Gibbs sampler for Bayesian inference in our LADY network model. Given P\`olya-gamma augmented data, our model can be recast as a Gaussian state space model for transformed data. By block-sampling in turn the latent coordinate processes for each actor $v$ conditionally on the latent positions of the others $u=1, \ldots, V, \ u \neq v$, we obtain a linear observation equation, which allows us to apply standard results from Kalman filtering \citep{durb_2012}. Posterior computation alternates between the following steps:
\begin{description}
\item{{\bf Step [1]}: Update  augmented data $\omega_{vu}(t_i)$ from the full conditional P\'olya-gamma,
$ \omega_{vu}(t_i) \mid - \sim \mbox{\small{PG}}\{1,\mu(t_i)+\sum_{h=1}^{H} x_{vh}(t_i)x_{uh}(t_i)\},$
for each $v=2,\ldots,V$, $u=1,\ldots,v-1$ and time $t_i=t_1, \ldots, t_n$.}

\vspace{7pt}

\item{{\bf Step [2]}: Adapting \eqref{eq5} to our model, the likelihood  for $\mu=\{ \mu(t_1), \ldots \mu(t_n)\}^{\T}$ given the P\'olya-gamma augmented data and the latent coordinate processes is
\begin{eqnarray*}
&\propto&\prod_{i=1}^n \exp \left[-\sum_{[vu]: v>u}\frac{\omega_{vu}(t_i)}{2}\left\{(Y_{t_i[vu]}-0.5)/\omega_{vu}(t_i)-\mu(t_i)-x_{v}(t_i)^{\T}x_{u}(t_i)\right\}^2 \right],\\
&\propto&\prod_{i=1}^n \exp \left[-\frac{\sum_{[vu]: v>u}\omega_{vu}(t_i)}{2}\left\{\mu(t_i)^2-2\mu(t_i) \frac{\sum_{[vu]: v>u}r_{vu}(t_i)}{\sum_{[vu]: v>u}\omega_{vu}(t_i)}\right\} \right],\\
&\propto&\prod_{i=1}^n \exp \left[-\frac{\sum_{[vu]: v>u}\omega_{vu}(t_i)}{2}\left\{ \frac{\sum_{[vu]: v>u}r_{vu}(t_i)}{\sum_{[vu]: v>u}\omega_{vu}(t_i)}-\mu(t_i)\right\}^2 \right],
\end{eqnarray*}
with $r_{vu}(t_i)=Y_{t_i[vu]}-0.5-\omega_{vu}(t_i)x_{v}(t_i)^{\T}x_{u}(t_i)$. Let $\omega_{\mu(t_i)}=\sum_{[vu]: v>u}\omega_{vu}(t_i)\in \Re^+$  and $Y_{\mu(t_i)}=\sum_{[vu]: v>u}r_{vu}(t_i)/\sum_{[vu]: v>u}\omega_{vu}(t_i)\in \Re$ for $i=1, \ldots, n$, the above likelihood for the baseline vector $\mu$ arises from the model
\begin{eqnarray}
Y_{\mu(t_i)}=\mu(t_i)+\epsilon_{\mu(t_i)}, \quad \mbox{independently for every } i=1, \ldots, n, 
\label{obs_mu}
\end{eqnarray}
with $\epsilon_{\mu(t_i)} \sim \mbox{N}(0, 1/\omega_{\mu(t_i)})$. Hence, the observation equation \eqref{obs_mu}  and the state equations  \eqref{eq:3}, define a Gaussian linear state space model, which allows updating for $\mu=\{ \mu(t_1), \ldots \mu(t_n)\}^{\T}$, $\mu'=\{ \mu'(t_1), \ldots \mu'(t_n)\}^{\T}$ and $z=\{ z(t_1), \ldots z(t_n)\}^{\T}$ via the simulation smoother of \citet{durb_2002}.  This has a computational complexity of $O(n)$ and diffuse initialization at $t_1$, $\{\mu(t_1),\mu'(t_1),z(t_1) \}^{\T} \sim \mbox{N}_3\{0,\mbox{diag}(100,100,100)\}$.}

\vspace{7pt}

\item{{\bf Step [3]}: For every actor $v=1, \ldots, V$, we rely on similar derivations to block-sample the latent coordinate trajectories $\{x_{vh}(t_i): h=1, \ldots, H, \ t_i=t_1, \ldots, t_n \}$, along with their first derivatives $\{x'_{vh}(t_i): h=1, \ldots, H, \ t_i=t_1, \ldots, t_n \}$ and the local instantaneous means $\{m_{vh}(t_i): h=1, \ldots, H, \ t_i=t_1, \ldots, t_n \}$. Specifically, given the baseline process, the P\`olya-gamma augmented data, and the remaining row processes $\{x_{uh}(t_i): u \neq v, \ h=1, \ldots, H, \ t_i=t_1, \ldots, t_n \}$, we obtain the following linear observation equations in $x_{v}(t_i)$
\begin{eqnarray}
\ \ \ \ \ \ \ \ \ \ \ \ \ {Y}_{x_v(t_i)}=X_{(-v)}(t_i)x_{v}(t_i)+\epsilon_{x_v(t_i)}, \quad \mbox{independently for every } i=1, \ldots, n, 
\label{obs_x}
\end{eqnarray}
where $X_{(-v)}(t_i)$ is the $(V-1) \times H$ coordinate matrix at time $t_i$ with the $v$th row held out, $Y_{x_v(t_i)}$ denotes the $(V-1) \times 1$ vector of transformed data $Y_{x_v(t_i)}= \mbox{diag}\{\Omega_{(v)}(t_i)\}^{-1} \{Y_{t_i(v)}-0.5 \cdot \mbox{1}_{V-1}-\mu(t_i)\Omega_{(v)}(t_i)\}$ and $\epsilon_{x_v(t_i)}$ is the noise vector $\epsilon_{x_v(t_i)} \sim \mbox{N}_{V-1}(0, \mbox{diag}\{\Omega_{(v)}(t_i)\}^{-1})$ for $i=1, \ldots, n$. In the above notation $\Omega_{(v)}(t_i)$ is the $(V-1) \times 1$ vector containing the P\'olya-gamma augmented data $\omega_{vu}(t_i)$ at time $t_i$ corresponding to all the pairs of actors having $v$ as one of the two. The same holds for $Y_{t_i(v)}$. As in step [2], the observation equation \eqref{obs_x} along with state equations in \eqref{eq:2} for $v$ and $h=1, \ldots, H$ form a linear Gaussian state space model from which the processes $x_{vh}=\{ x_{vh}(t_1), \ldots x_{vh}(t_n)\}^{\T}$, $x_{vh}'=\{ x_{vh}'(t_1), \ldots x_{vh}'(t_n)\}^{\T}$ and $m_{vh}=\{ m_{vh}(t_1), \ldots m_{vh}(t_n)\}^{\T}$ can be updated via simulation smoothing  \citep{durb_2002} under the same diffuse initialization at $t_1$, $\{x_{vh}(t_1),x_{vh}'(t_1),m_{vh}(t_1) \}^{\T} \sim \mbox{N}_3\{0,\mbox{diag}(100,100,100)\}$ for each $h=1, \ldots, H$.}

\vspace{7pt}

\item{{\bf Step [4]}: Letting $\sigma_{\mu}^2 \sim \mbox{Inv-Ga}(a_{\mu},b_{\mu})$ and $\sigma_{z}^2 \sim \mbox{Inv-Ga}(a_{z},b_{z})$, the hyper-priors for the noise variances in the states equations   \eqref{eq:3}, their full conditionals  are
\begin{eqnarray*}
\sigma_{\mu}^{2}\mid- &\sim& \mbox{Inv-Ga} \left[ a_{\mu}+\frac{n-1}{2},b_{\mu}+\frac{1}{2} \sum^{n-1}_{i=1} \frac{\{\mu'(t_{i+1}) - \mu'(t_{i})-z(t_{i})\delta_{i}\}^{2}}{\delta_{i}} \right], \nonumber \\
\sigma_{z}^{2}\mid- &\sim& \mbox{Inv-Ga} \left[ a_{z}+\frac{n-1}{2},b_{z}+\frac{1}{2} \sum^{n-1}_{i=1} \frac{\{z(t_{i+1})-z(t_{i})\}^{2}}{\delta_{i}} \right].
\end{eqnarray*}}

\item{{\bf Step [5]}: Similarly, assuming the variances  $\sigma_{x_{vh}}^2 \sim \mbox{Inv-Ga}(a_{x},b_{x})$ and $\sigma_{m_{vh}}^2 \sim \mbox{Inv-Ga}(a_{m},b_{m})$, independently for each $v=1, \ldots,V$ and $h=1, \ldots, H$, their full conditionals are
\begin{eqnarray*}
\sigma_{x_{vh}}^2\mid- &\sim& \mbox{Inv-Ga} \left[ a_{x}+\frac{n-1}{2},b_{x}+\frac{1}{2} \sum^{n-1}_{i=1} \frac{\{x_{vh}'(t_{i+1}) - x_{vh}'(t_{i})-m_{vh}(t_{i})\delta_{i}\}^{2}}{\delta_{i}} \right], \nonumber \\
\sigma_{m_{vh}}^2\mid- &\sim& \mbox{Inv-Ga} \left[ a_{m}+\frac{n-1}{2},b_{m}+\frac{1}{2} \sum^{n-1}_{i=1} \frac{\{m_{vh}(t_{i+1})-m_{vh}(t_{i})\}^{2}}{\delta_{i}} \right],
\end{eqnarray*}
for each $v=1, \ldots, V$ and $h=1, \ldots, H$.}

\vspace{7pt}

\item{{\bf Step [6]}: Finally, given the posterior samples for the baseline process  $\mu=\{ \mu(t_1), \ldots \mu(t_n)\}^{\T}$ and $x_{vh}=\{ x_{vh}(t_1), \ldots x_{vh}(t_n)\}^{\T}$, for each $v=1, \ldots, V$ and $h=1, \ldots H$, obtain the posterior samples for the dynamic edge probabilities by applying \eqref{eq1.1} as follow
\begin{eqnarray*}
\pi_{vu}(t_i)=\big[1+\exp\{-\mu(t_i)-x_{v}(t_i)^{\T}x_{u}(t_i)\}\big]^{-1},
\end{eqnarray*}
for each $v=2, \ldots, V$, $u=1,\ldots,v-1$ and time $t_i=t_1,\ldots, t_n$.}
\end{description}

To choose $H$, we repeat the above algorithm for increasing $H$, stopping when there is no substantial improvement for in-sample edge prediction based on the area under the ROC curve (AUC). As in-sample predictive strategies may suffer from over-fitting issues, we additionally assess our choice of $H$ by exploring out-of-sample prediction and forecasting performance.

\subsection{Forecasting and predicting}\label{fore}
Forecasting a future network based on past data is particularly appealing for our motivating application in facilitating the design of specific policies; for example, aimed at outbreak prevention.  If an individual contracts a disease at time $t_n$, forecasts at time $t_{n+1}$ are a key to understand which students are at risk of contagion as a result of face-to-face proximity interactions. 

Under our Bayesian paradigm, a strategy to obtain one-step-ahead forecasts of future edges is to rely on the expectation of the forecasted predictive distribution defined as
\begin{eqnarray}
\mbox{E}\{\mathcal{Y}_{vu}(t_{n+1}) \mid Y_{t_1}, \ldots, Y_{t_n}\}&=&\mbox{E}_{\pi_{vu}(t_{n+1})}[\mbox{E}_{\mathcal{Y}_{vu}(t_{n+1})}\{\mathcal{Y}_{vu}(t_{n+1}) \mid \pi_{vu}(t_{n+1})\} \mid Y_{t_1}, \ldots, Y_{t_n}] \nonumber\\
&=&\mbox{E}[\pi_{vu}(t_{n+1}) \mid Y_{t_1}, \ldots, Y_{t_n}],
\label{for}
\end{eqnarray}
for each $v=2, \ldots, V$ and $u=1, \ldots, v-1$. Hence equation \eqref{for} simply requires the posterior mean of the edge probabilities at time $t_{n+1}$.  The Markovian property implied by our state equations in \eqref{eq:2}--\eqref{eq:3} provides a natural procedure to obtain these quantities along with the entire posterior distribution for $\pi_{vu}(t_{n+1})$, for each $v=2, \ldots, V$, $u=1, \ldots, v-1$. Specifically, according to  \eqref{eq1.1}--\eqref{eq:3} samples from the posterior of $\pi_{vu}(t_{n+1})$ can be simply obtained by applying the equation
\begin{eqnarray*}
\pi_{vu}(t_{n+1})=\left(1+\exp[-\{\mu(t_n)+\delta_{n}\mu'(t_n)\}-\{x_{v}(t_n)+\delta_{n}x_{v}'(t_n)\}^{\T}\{x_{u}(t_n)+\delta_{n}x_{u}'(t_n)\}]\right)^{-1},
\label{pi_for}
\end{eqnarray*}
to the posterior samples of the latent states at time $t_n$, for $v=2, \ldots, V$ and $u=1,\ldots,v-1$.

Recalling our data set structure, beside forecasting contacts at the next time within the first day, it is additionally of interest to predict the whole network dynamics in the second day, based on the estimates from the previous day. In particular, letting $\mathcal{Y}^*(t_i)$ denote the random matrix encoding presence or absence of contacts among pairs of actors at time $t_i$ in the second day, we predict the edges in $\mathcal{Y}^*(t_i)$ by focusing on 
\begin{eqnarray}
\mbox{E}\{\mathcal{Y}^*_{vu}(t_i) \mid Y_{t_1}, \ldots, Y_{t_n}\}&=&\mbox{E}_{\pi_{vu}(t_{i})}[\mbox{E}_{\mathcal{Y}^*_{vu}(t_i)}\{\mathcal{Y}^*_{vu}(t_i) \mid \pi_{vu}(t_{i})\} \mid Y_{t_1}, \ldots, Y_{t_n}] \nonumber\\
&=&\mbox{E}[\pi_{vu}(t_{i}) \mid Y_{t_1}, \ldots, Y_{t_n}],
\label{pred}
\end{eqnarray}
for each $v=2, \ldots, V$, $u=1, \ldots, v-1$ and time $t_i$, where the expectation in \eqref{pred} coincides with the posterior mean of the edge probability trajectories. Equation \eqref{pred} relies on the assumption that the dynamic contacts at the second day are governed by the same process underlying data in the first day.  If we had data on multiple days, we could refine our model to include dynamic changes across days instead of just within a given day,  but we avoid such complexity here and use the second day as a test set to evaluate predictive performance.

\subsection{Online updating} \label{online}
Once the model has been estimated on data $Y_{t_1}, \ldots, Y_{t_n}$, new contact networks $Y_{t_{n+1}}, \ldots, Y_{t_{n+\bar{n}}}$ can stream in. In order to rapidly update policies, such as disease surveillance, it is important to  have a fast online updating algorithm for the posterior of the edge probabilities $\pi_{vu}(t_{n+1}), \ldots, \pi_{vu}(t_{n+\bar{n}})$, $v=2, \ldots, V$, $u=1, \ldots, v-1$, including information from new networks $Y_{t_{n+1}}, \ldots, Y_{t_{n+\bar{n}}}$,  without the need to rerun posterior computation for the whole data from $t_1$ to $t_{n+\bar{n}}$.

Our LADY network model is amenable to fast updating due to the latent Kalman filter formulation. Exploiting the posterior means and covariances of the latent states at time $t_{n}$ and the estimated noise variances in the state equation, our online updating algorithm efficiently cycles between steps [1]--[3] and [6] of the Gibbs sampler only for new data $Y_{t_{n+1}}, \ldots Y_{t_{n+\bar{n}}}$, with the simulation smoother in [2] and [3] initialized at $t_{n+1}$ using the predictive distribution from the Kalman filter. Specifically we initialize states $\{\mu(t_{n+1}),\mu'(t_{n+1}),z(t_{n+1}) \}^{\T}$ at $t_{n+1}$ in [2] by assuming  $\{\mu(t_{n+1}),\mu'(t_{n+1}),z(t_{n+1}) \}^{\T}$ are distributed according to
{{\begin{eqnarray*}
\mbox{N}_3\{T_n[\hat{\mbox{E}}\{{\mu}(t_n)\}, \hat{\mbox{E}}\{{\mu'}(t_n)\}, \hat{\mbox{E}}\{{z}(t_n)\}]^{\T}, T_n\hat{\Gamma}_{\mu,n} T_n^{\T}+R_n \mbox{diag}(\hat{\sigma}^2_{\mu} \delta_{n}, \hat{\sigma}^2_{z} \delta_{n}) R_n^{\T}\},
\end{eqnarray*}}}where $[\hat{\mbox{E}}\{{\mu}(t_n)\}, \hat{\mbox{E}}\{{\mu'}(t_n)\}, \hat{\mbox{E}}\{{z'}(t_n)\}]^{\T}$ is the vector of posterior means for the latent states at time $t_n$, $\hat{\Gamma}_{\mu,n}$ is their $3 \times 3$ posterior covariance matrix and $\hat{\sigma}^2_{\mu}$, $\hat{\sigma}^2_{z}$ are the estimated noise variances using the initial data set from $t_1$ to $t_n$. A similar initialization is considered in step [3] for $\{x_{vh}(t_{n+1}),x_{vh}'(t_{n+1}),m_{vh}(t_{n+1}) \}^{\T}$ obtaining
{{\begin{eqnarray*}
\mbox{N}_3\{T_n[\hat{\mbox{E}}\{{x_{vh}}(t_n)\}, \hat{\mbox{E}}\{{x_{vh}'}(t_n)\}, \hat{\mbox{E}}\{{m_{vh}}(t_n)\}]^{\T}, T_n\hat{\Gamma}_{x_{vh},n} T_n^{\T}+R_n \mbox{diag}(\hat{\sigma}^2_{x_{vh}} \delta_{n}, \hat{\sigma}^2_{m_{vh}} \delta_{n}) R_n^{\T}\},
\end{eqnarray*}}}for $v=1, \ldots, V$ and $h=1, \ldots, H$.

Although the algorithm fixes the hyperparameters corresponding to the noise variances in the state equations at their posterior means, these quantities are time-constant and hence can be accurately estimated by borrowing information across the whole time window. It is however straightforward to modify the algorithm to update the posterior distribution also for these quantities given the latent states stored in the initial sampling from $t_1$ to $t_n$ and the updated ones from $t_{n+1}$ to $t_{n+\bar{n}}$. This strategy may be useful when $n$ is small. We found few differences between the two procedures in our simulations and hence prefer the first strategy. 

It is also worth noticing that our procedure does not update the previous $\pi_{vu}(t_{1}), \ldots, \pi_{vu}(t_{n})$, $v=2, \ldots, V$, $u=1, \ldots, v-1$ given new data $Y_{t_{n+1}}, \ldots Y_{t_{n+\bar{n}}}$, but focuses only on the posterior of $\pi_{vu}(t_{n+1}), \ldots, \pi_{vu}(t_{n+\bar{n}})$, $v=2, \ldots, V$, $u=1, \ldots, v-1$. This may affect the ability of our procedures to properly propagate uncertainty and reduce performance in updating $\pi_{vu}(t_{n+1}), \ldots, \pi_{vu}(t_{n+\bar{n}})$, $v=2, \ldots, V$, $u=1, \ldots, v-1$. To mitigate this issue, while maintaing computational scalability, we run online updating for data $Y_{t_{n-j}}, \ldots, Y_{t_n},Y_{t_{n+1}}, \ldots Y_{t_{n+\bar{n}}}$ instead of only $Y_{t_{n+1}}, \ldots Y_{t_{n+\bar{n}}}$. We found this correction to improve performance even when a small number $j$ of past networks is included along with new data.

\subsection{Model checking}\label{check}
Our LADY network formulation and related procedures fall within the class of latent variable modeling of dynamic networks. Although these methodologies are appealing in accommodating heterogenous structures and facilitate tractable inference, the types of higher-order dependencies included may be limited by the conditional independence assumption and the characterization of the latent variables. Although conditional independence may at first appear overly-restrictive, multivariate categorical data --- such as a vector of edges --- can be expressed as conditionally independent given a sufficient number of latent factors without imposing any assumptions on the joint distribution; see for example \citet{dun_2009} for theoretical results. 

To check the flexibility of our model, we develop approaches for assessing model adequacy.  In the Bayesian literature, it is common to rely on diagnostics comparing the posterior predictive distribution associated with the model to the observed data; refer to  \cite{gel_2014} for an overview. In our case the posterior predictive distribution is defined as
\begin{eqnarray*}
\mbox{pr}\{\mathcal{Y}(t_1), \ldots, \mathcal{Y}(t_n) \mid Y_{t_1}, \ldots, Y_{t_n}\}=\int  \prod_{i=1}^n \prod_{[vu]: v>u}  \mbox{pr}\{\mathcal{Y}_{vu}(t_i) \mid \pi_{vu}(t_i)\} \mbox{d}\Pi\{\pi \mid Y_{t_1}, \ldots, Y_{t_n}\}, 
\label{post_pred}
\end{eqnarray*}
where $ \mbox{pr}\{\mathcal{Y}_{vu}(t_i) \mid \pi_{vu}(t_i)\}$ is the Bernoulli probability mass function in \eqref{eq1}, while the quantity $\Pi\{\pi \mid Y_{t_1}, \ldots, Y_{t_n}\}$ denotes the joint posterior distribution for the trajectories of the edge probabilities given the observed data $Y_{t_1}, \ldots, Y_{t_n}$. It is straightforward to simulate from $\mbox{pr}\{\mathcal{Y}(t_1), \ldots, \mathcal{Y}(t_n) \mid Y_{t_1}, \ldots, Y_{t_n}\}$ exploiting equation \eqref{eq1} along with posterior samples for $ \pi_{vu}(t_i)$, $v=2, \ldots, V$, $u=1, \ldots, v-1$ and $t_i=t_1, \ldots, t_n$. Specifically, for each MCMC sample of  $ \pi_{vu}(t_i)$, we simulate contacts among the corresponding pair of actors from conditionally independent Bernoulli random variables given $ \pi_{vu}(t_i)$. 

Exploiting the samples from the posterior predictive distribution, we evaluate the performance of our model in accommodating specific topological structures observed from the data. We focus on the dynamic network density $\sum_{[vu]:v>u} \mathcal{Y}_{vu}(t_i)/\{V(V-1)/2\}$, the time-varying actor degree $\sum_{u\neq v} \mathcal{Y}_{vu}(t_i)$, for $v=1, \ldots, V$, and  the dynamic homophily by class and gender measured by the assortativity coefficient; see \citet{new_2003}, equation 2.

When the interest is on disease surveillance and outbreak prevention, the dynamic network density is a key quantity in summarizing the frequency of contacts including those leading to potential contagion. Actors' degrees are appealing in providing a measure of the number of subjects at risk of contagion if an individual contracts a disease at a certain time. Evolution of homophily structures across time and environmental conditions are of interest from a social science perspective; see for example  \citet{ste_2013} for a study on gender homophily in face-to-face contact networks from an averaged perspective. In assessing model adequacy, we compare these network summary measures computed from the observed data to the posterior predictive distribution of these quantities generated under the presumed model.  If the model is not sufficiently flexible, we expect the observed measures to fall in the tails of their posterior predictive distributions.  We perform also out-of-sample assessments to evaluate overall performance of the model in characterizing the data.

\section{Simulation study} \label{sim}
\begin{figure}
\centering
\includegraphics[trim=1cm 0.5cm 0.5cm 0cm, clip=true,height=7cm, width=15.5cm]{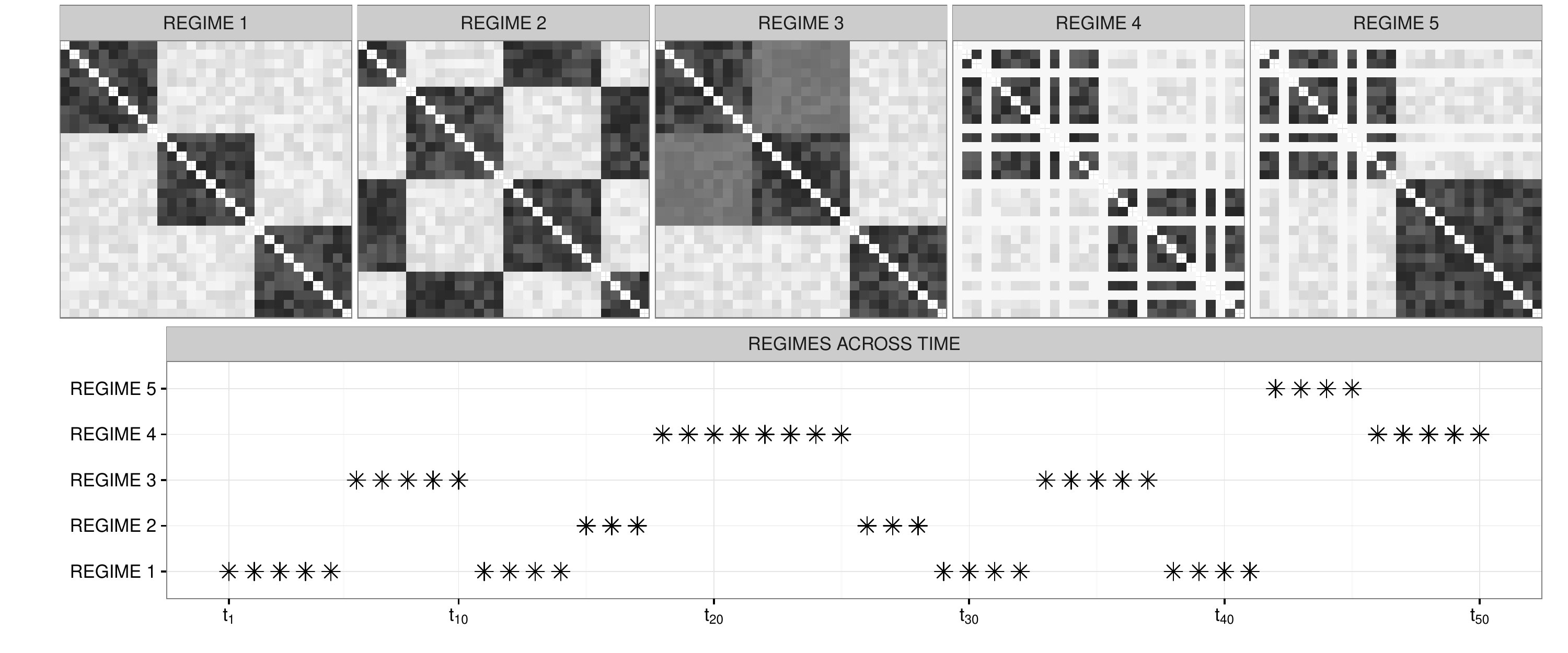}
\caption{\footnotesize{Upper panels: true edge probabilities --- arranged in matrix form --- for the regimes in the simulation; colors go from white to black as the probability goes from 0 to 1. Lower panels: graphical representation showing for every time which regime --- i.e. true edge probabilities --- is considered to simulate the data.}}
\label{Sim1}
\end{figure}

We implement a simulation study to assess the performance of our LADY network model in estimating trajectories with varying smoothness, accommodating streaming data and predicting future networks. We consider dynamic networks with $V=30$ actors monitored for $n=50$ equally spaced times from $t_1=0$ to $t_{50}=15$. The time varying edges $Y_{t_i[vu]}$ are simulated from model \eqref{eq1} with edge probabilities evolving in time across five regimes mimicking --- in a simple version --- possible scenarios associated with our face-to-face student interactions. Refer to Figure \ref{Sim1} for a representation of the true edge probabilities.

Specifically we consider three classes comprising ten students each and define also a gender variable. There are $5$ males and $5$ females  in each class, corresponding to the subsets of actors $V_{\mbox{{\footnotesize{m}}}} =\{1, \ldots, 5, 16, \ldots, 25 \}$ and $V_{\mbox{{\footnotesize{f}}}}=\{6, \ldots, 15, 26, \ldots, 30\}$, respectively. The first regime represents school hours and is characterized by high probability of contact between students in the same class, and low chance of face-to-face interaction among students in different classes. The second regime encodes high gender homophily, which may arise during the breaks before and after lunch times when all the students can interact; see also analyses in \citet{ste_2013}.  The third regime is characterized by the first two classes sharing the same room --- for school hours or breaks --- and hence, beside high within class probabilities of contact, we observe also a moderately high chance of contact between students in the first two classes. Regime four represents a possible scenario we have observed in our data during lunch times and confirmed in Figure 10 of  \citet{steh_2011}. Specifically, students in the second class are equally divided in two groups, with one attending lunch with students in the first class and the other with those in the third class. Hence we observe two block structures, with an additional subset of the students having no contacts with the others in leaving the school for lunch. Regimes five and four define also networks during the end of the school day, with groups of students gathering in the same room and progressively leaving the school. Actors $1$, $4$, $10$ and $12$ go home at time $t_{42}$, whereas actors $16$, $20$, $26$ and $28$ leave the school at $t_{46}$.

Although this generative mechanism represents a substantially simplified version of our complex data set, the basic underlying structures and the rapid changes in specific topological patterns are in line with those we expect in our application. Moreover, considering edge probabilities obtained under representations different than \eqref{eq1.1} and evolving in time across a regime-switching process instead of the  state equations \eqref{eq:2}--\eqref{eq:3} has the additional benefit of providing a more effective validation of our LADY network methodology, as the true edge probability processes are not generated from our model.

\subsection{Posterior inference and model checking}
In performing posterior inference, we choose moderately diffuse priors for the noise variances in the state equations by letting $a_{\mu}=a_{z}=a_{x}=a_{m}=b_{\mu}=b_{z}=b_{x}=b_{m}=0.01$, and  run $5{,}000$ Gibbs iterations discarding the first $1{,}000$. To learn $H$ we consider our selection procedure by performing posterior computation for increasing $H=1,2, \ldots$ and provide posterior inference for the model having $H$ total latent coordinates such that $\mbox{AUC}_{H+1} -\mbox{AUC}_{H}<0.01$. The AUC for the model with only the baseline process is $0.603$, while those for formulations with $H=1$ and $H=2$ are $0.901$ and $0.943$, respectively. Increasing the coordinates from $H=2$ to $H=3$ we found no substantial improvement with an AUC of $0.947$, so $H=2$ is chosen.

\begin{figure}
\centering
\includegraphics[trim=0.7cm 0.5cm 0.2cm 0.2cm, clip=true,height=7cm, width=15.5cm]{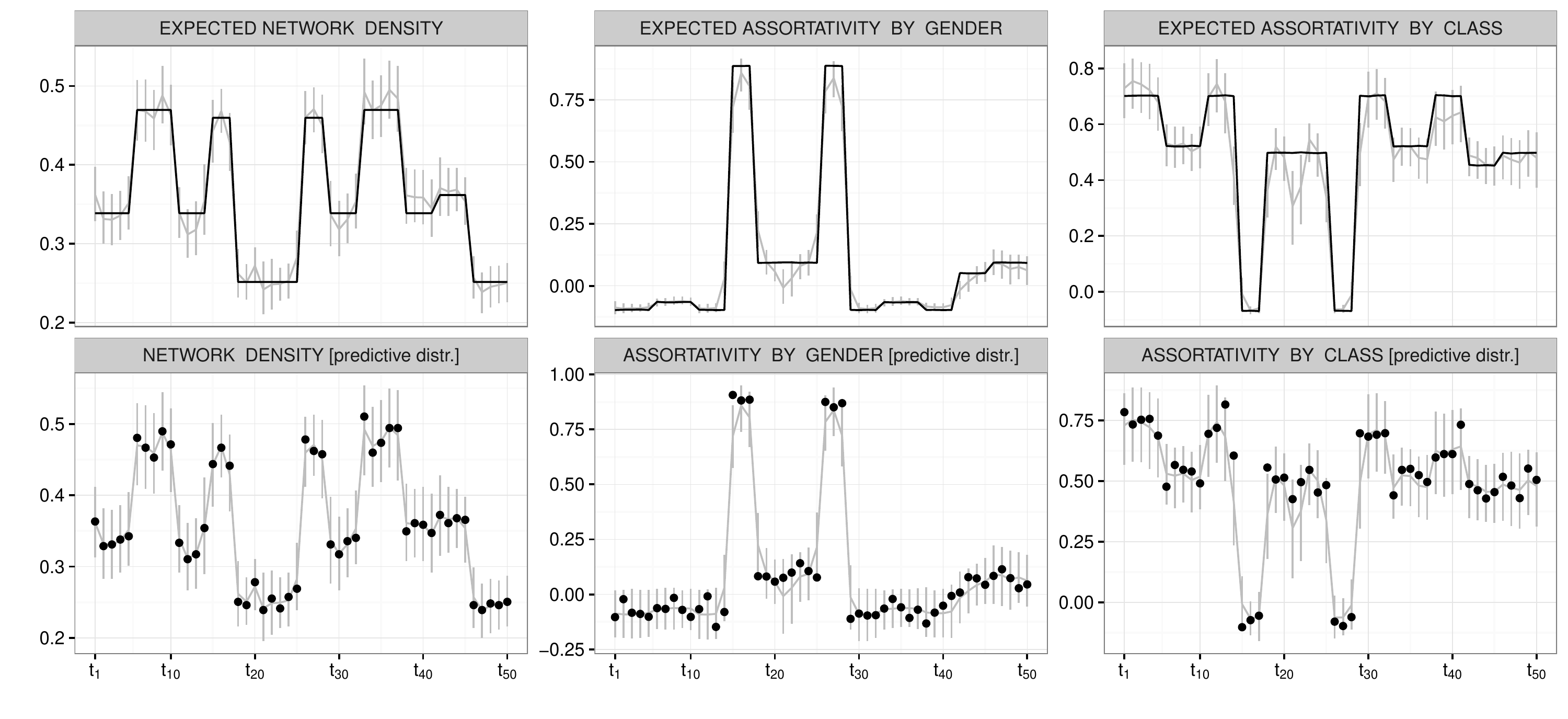}
\caption{\footnotesize{Upper panels: trajectory of the posterior mean (grey line) and point-wise $0.95$ highest posterior density intervals (grey segments) for dynamic expected network summary measures covering network density, assortativity by gender and by class; true trajectories are represented by the black line. Bottom panels: for the same summary measures, mean trajectory (grey line) and point-wise $0.95$ predictive intervals (grey segments) obtained from the posterior predictive distribution; black dots represent the corresponding time-varying network measures computed from the simulated data.}}
\label{Sim2}
\end{figure}

Convergence is assessed by visual inspection of the traceplots for quantities of interest, and by the \citet{gel_1992} potential scale reduction factors (PSRFs). These quantities are obtained by comparing between and within sub-chains variances, after splitting each chain of interest in four consecutive sub-chains of length $1{,}000$, after burn-in. The median of the PSRFs for the chains of the edge probabilities $\pi_{vu}(t_i)$, $v=2, \ldots, V$, $u=1,\ldots, v-1$ and $t_i=t_1, \ldots, t_n$, is $1.01$, with $99\%$ being less than $1.15$, providing evidence of convergence. Similar results are obtained for the network measures of interest, including the dynamic expected density  $\mbox{E}[\sum_{[vu]:v>u} \mathcal{Y}_{vu}(t_i)/\{V(V-1)/2\}]=\sum_{[vu]:v>u} \mbox{E}\{ \mathcal{Y}_{vu}(t_i)\}/\{V(V-1)/2\}=\sum_{[vu]:v>u}\pi_{vu}(t_i)/\{V(V-1)/2\}$, the time-varying expected actor degree $\mbox{E}\{ \sum_{u\neq v} \mathcal{Y}_{vu}(t_i)\}= \sum_{u\neq v}\mbox{E}\{ \mathcal{Y}_{vu}(t_i)\}= \sum_{u\neq v} \pi_{vu}(t_i)$ for each $v=1, \ldots, V$, and  the dynamic expected homophily by class and gender. As the expectation of the assortativity coefficients is not analytically available as a function of the edge probabilities, we obtain posterior samples for the expected assortativity coefficients via Monte Carlo methods. Specifically, for each posterior sample of $\pi_{vu}(t_i)$, $v=2, \ldots, V$, $u=1,\ldots, v-1$ and $t_i=t_1, \ldots, t_n$, we simulate 100 networks from \eqref{eq1} and obtain approximated samples from the posterior distribution of the dynamic expected assortativity by class and gender, by computing these coefficients for the 100 simulated networks and averaging.

\begin{figure}
\centering
\includegraphics[trim=0.8cm 0.6cm 0.1cm 0.1cm, clip=true,height=7cm, width=15.5cm]{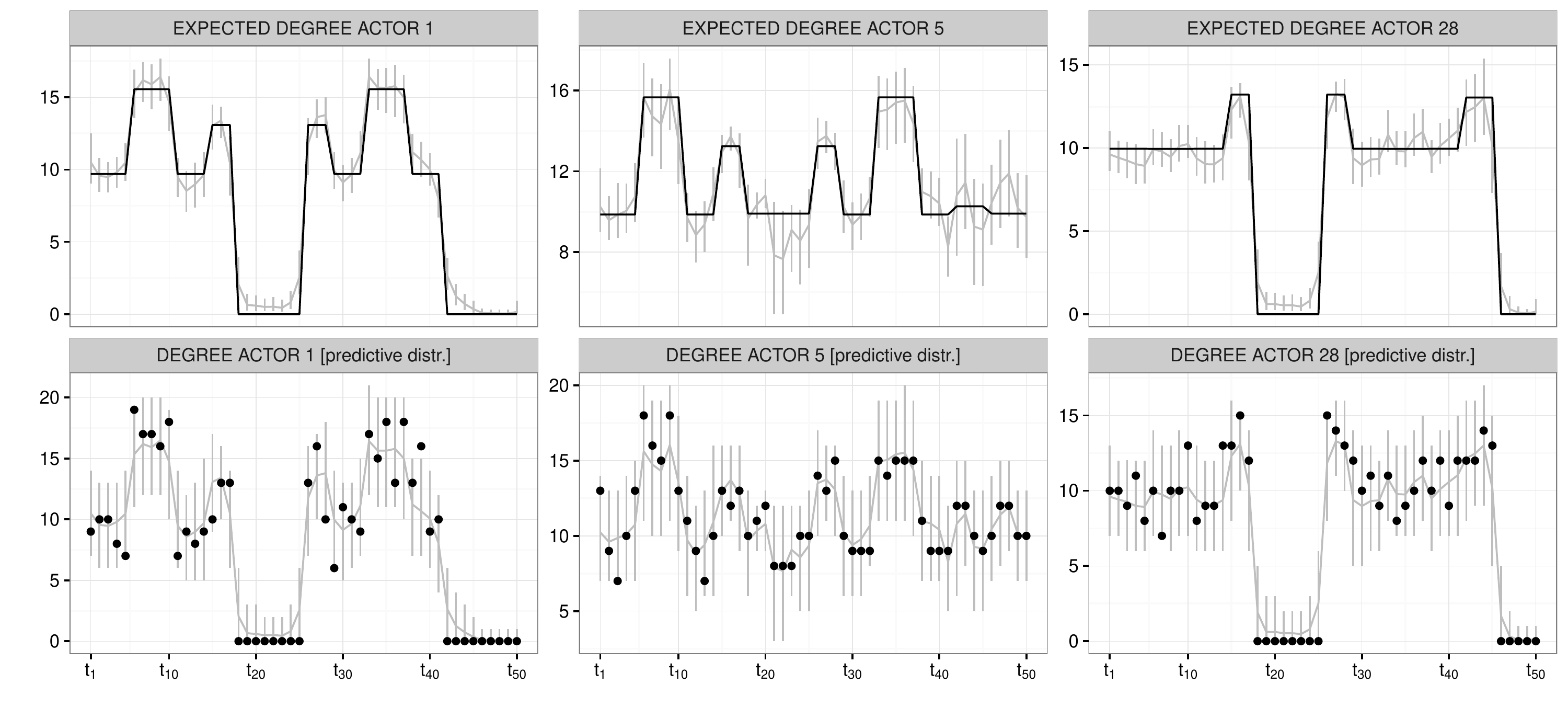}
\caption{\footnotesize{Upper panels: trajectory of the posterior mean (grey line) and point-wise $0.95$ highest posterior density intervals (grey segments) for the dynamic expected degree of selected actors; true trajectories are represented by the black line. Bottom panels: for the same summary measures, mean trajectory (grey line) and point-wise $0.95$ predictive intervals (grey segments) obtained from the posterior predictive distribution; black dots represent the corresponding time-varying actors' degrees computed from the simulated data.}}
\label{Sim3}
\end{figure}

As shown in the upper panels of Figures \ref{Sim2} and \ref{Sim3}, enforcing local adaptivity in the time-varying trajectories of the edge probabilities while accommodating across-actor heterogeneity, allows our model to capture rapid changes in the true expected measures of interest, including time-varying network density, homophily structures and actors' degrees. Moreover, although we rely on a latent space representation which does not explicitly parameterize dependencies among edges, our model can accurately accommodate topological structures of interest characterizing the observed dynamic networks. This is highlighted in the bottom panels  of Figures \ref{Sim2} and \ref{Sim3}, comparing network summary measures computed from the observed data with their  posterior predictive distribution --- consistent with the  procedures outlined in Section  \ref{check}. Almost all the observed quantities are inside the $0.95$ posterior predictive intervals.

\subsection{Online updating, forecasting and predictive performance}
Table \ref{tab:1} compares forecasting and predictive performance of our model to those associated with two selected competitors, for times from $t_{45}$ to $t_{50}$. Specifically, we consider the Gaussian process dynamic network model developed by \citet{dur_2014} and the temporal ERGM (TERGM) proposed by \citet{hann_2007}. \citet{dur_2014} rely on our model formulation \eqref{eq1}--\eqref{eq1.1} but do not allow varying smoothness over time. \citet{hann_2007} TERGM is instead a substantially different model which explicitly accounts for the effect of topological structures in the model formulation, rather than considering latent variables.

In performing posterior computation under \citealp{dur_2014}, we consider the same hyperparameter settings in their simulation study, fixing $H=2$ --- as in the LADY network model for this simulation --- with the GP length scales $\kappa_{\mu}=\kappa_{x}=0.01$.  Moderate changes in the length scales provided comparable results. The TERGM is instead estimated via bootstrapped pseudolikelihood procedures \citep{desm_2012} exploiting the \texttt{R} packages \texttt{btergm} and \texttt{xergm}. In defining the linear predictor under the TERGM representation we consider a $p^*$ ERGM specification with alternating $k$-stars \citep{robin_2007} and triangle effects to account for transitivity patterns. We additionally include gender and class variables via main and homophily effects --- using functions \texttt{nodefactor()} and \texttt{nodematch()}, respectively. Finally, we account for temporal dependence by including a stability term which measures the tendency of an edge --- or non-edge --- at time $t_i$ to be also observed --- or not observed --- at the next time $t_{i+1}$. The main effects of the actors' covariates were not significant, hence we drop these predictors in assessing forecasting and predictive performance. We additionally hold out the triangle effect, as the inclusion of this term substantially reduced forecasting and predictive performance. We also attempted an actor-oriented model using the \texttt{R} package \texttt{RSiena}  but found convergence issues for the parameters in the rate function.

\begin{table}[t]
		\centering
\caption{\footnotesize{For our model and the two competitors, forecasting and predictive performance for times from $t_{45}$ to $t_{50}$.}}
{\renewcommand{\arraystretch}{1.3}%

			\begin{tabular}{lllllllll}

&$t_{45}$ & $t_{46}$&$t_{47}$&$t_{48}$&$t_{49}$&$t_{50}$\\
		\hline
&\multicolumn{6}{c}{ {\bf Forecasting Performance} monitored via areas under the ROC curve}\\
\hline
LADY& {\scriptsize{0.913}} {\tiny{[0.90,0.92]}}&{\scriptsize{0.837}} {\tiny{[0.83,0.85]}} &  {\scriptsize{0.916}} {\tiny{[0.91,0.92]}}& {\scriptsize{0.923}} {\tiny{[0.92,0.93]}}& {\scriptsize{0.936}} {\tiny{[0.93,0.95]}} & {\scriptsize{0.935}} {\tiny{[0.93,0.94]}}\\
GP  & {\scriptsize{0.894}} {\tiny{[0.88,0.91]}}& {\scriptsize{0.803}} {\tiny{[0.79,0.81]}} & {\scriptsize{0.891}} {\tiny{[0.88,0.90]}} &{\scriptsize{0.921}} {\tiny{[0.91,0.93]}} & {\scriptsize{0.916}} {\tiny{[0.90,0.93]}}&{\scriptsize{0.923}} {\tiny{[0.92,0.93]}} \\ 
TERGM &{\scriptsize{0.877}} {\tiny{[0.84,0.89]}}&{\scriptsize{0.739}} {\tiny{[0.57,0.84]}} &{\scriptsize{0.847}} {\tiny{[0.74,0.90]}}  &{\scriptsize{0.818}} {\tiny{[0.73,0.88]}} &  {\scriptsize{0.848}} {\tiny{[0.77,0.89]}}& {\scriptsize{0.855}} {\tiny{[0.79,0.91]}} \\
		\hline
&\multicolumn{6}{c}{ {\bf Predictive Performance} monitored via areas under the ROC curve}\\
\hline
LADY& {\scriptsize{0.958}} {\tiny{[0.95,0.97]}}& {\scriptsize{0.958}} {\tiny{[0.95,0.96]}}& {\scriptsize{0.956}} {\tiny{[0.95,0.96]}} & {\scriptsize{0.954}} {\tiny{[0.95,0.96]}}&  {\scriptsize{0.949}} {\tiny{[0.94,0.96]}}& {\scriptsize{0.936}} {\tiny{[0.93,0.94]}}\\
GP &{\scriptsize{0.944}} {\tiny{[0.93,0.96]}}& {\scriptsize{0.940}} {\tiny{[0.93,0.95]}}& {\scriptsize{0.934}} {\tiny{[0.92,0.95]}} & {\scriptsize{0.923}} {\tiny{[0.91,0.94]}}&  {\scriptsize{0.933}} {\tiny{[0.92,0.94]}}& {\scriptsize{0.925}} {\tiny{[0.91,0.93]}} \\
TERGM&{\scriptsize{0.877}} {\tiny{[0.82,0.92]}}& {\scriptsize{0.873}} {\tiny{[0.81,0.90]}} & {\scriptsize{0.888}} {\tiny{[0.83,0.92]}} &{\scriptsize{0.877}} {\tiny{[0.83,0.91]}} & {\scriptsize{0.879}} {\tiny{[0.80,0.92]}}& {\scriptsize{0.891}} {\tiny{[0.87,0.90]}} \\
\hline

\end{tabular}}

\label{tab:1}

	\end{table}

For each time $t_i=t_{44},\ldots, t_{49}$ forecasting performance is assessed by estimating the three different models using data from $t_{1}$ to $t_{i}$, and forecasting edges at time $t_{i+1}$. Forecasts under the GP dynamic network follow procedures outlined in \citet{dur_2014}. Under the TERGM, forecasting of future networks proceeds via simulation methods using the  \texttt{gof()} function in the  \texttt{R} package  \texttt{ergm}; see also \citet{hunter_2008}. For our LADY network model, we  proceed by first updating the posterior distributions of the edge probabilities at $t_{i}$ using estimates from $t_{1}$ to $t_{i-1}$ according to procedures in Section \ref{online} --- with $j=5$ --- and then forecast edges at time $t_{i+1}$ by applying the forecasting methods in Section \ref{fore} to the posterior distributions of the edge probabilities from the online updating. Joining online updating and forecasting is appealing in providing a fast strategy which avoids re-running posterior computation for the whole data set when a one-step-ahead forecast in required.

In evaluating predictive performance we instead simulate new networks from the same mechanism considered to generate training data --- see Figure \ref{Sim1} --- and compare the areas under the ROC curves when predicting their edges  based on the estimates from the three competing methods --- exploiting training data $Y_{t_{1}}, \ldots, Y_{t_{50}}$. Edge prediction under our LADY network model and  \citet{dur_2014} GP dynamic network use equation \eqref{pred}. For TERGM we exploit again simulation procedures from the  \texttt{gof()} function. To more reliably assess  performance, we repeat the above forecasting and out-of-sample prediction exercises for $100$ different simulated data sets and report in Table \ref{tab:1} the median along with the $0.25$ and $0.75$ quantiles of the $100$ areas under the ROC curves obtained for every time.

As shown in Table \ref{tab:1} our procedure is characterized by improved forecasting and predictive performance compared to  the GP dynamic network model and TERGM.  \citet{dur_2014} accommodate heterogenous structures but assume time-constant smoothness.  \citet{hann_2007} explicitly account for higher-order dependencies but force the  model parameters to be shared among actors and constant across time. These assumptions lead to reduced performance compared to our procedure which incorporates both across-actor heterogeneity and time-varying smoothness. These results additionally highlight the good performance of our online updating procedures.

As expected, forecasting performance decreases at $t_{{46}}$ since the models have no experience of sudden regime changes. However, it is interesting to notice how incorporating local adaptivity provides rapid adjustments of the estimates to new regimes once they are observed, improving subsequent forecasts. The dynamic GP network model requires  more times to adapt to new regimes due to the time-constant smoothness assumption. Reduced performance at $t_{46}$ is not an issue when predicting new networks generated under the same mechanism, as the whole training data set $Y_{t_{1}}, \ldots, Y_{t_{50}}$ already inform on regime changes. In the out-of-sample prediction exercise, performance depends on the flexibility of the model in accommodating rapid regime changes along with their associated network structures.

Inference under our LADY network model takes $\sim 75$ minutes for posterior computation, $\sim 12$ minutes for online updating and $\sim 2$ second for forecasting. The dynamic GP network model requires  comparable time for forecasting but is  substantially slower in performing posterior computation --- $\sim 500$ minutes --- due to the computational bottlenecks of the Gaussian processes. Estimation under TERGM is faster, but simulation methods for forecasting and predictions require more time. Our algorithms are run in a naive \texttt{R} (version 3.2.1) implementation in a machine with one Intel Core i5 2.3GHz processor and 4GB of RAM. Hence, there are significant margins to further improve computational time.

\section{LADY networks for face-to-face interaction data}\label{app}
We apply our LADY network model outlined in Sections \ref{mod}--\ref{post} to the face-to-face contact data $Y_{t_1}, \ldots, Y_{t_{51}}$ described in Section \ref{data}, under the same settings of the simulation study, with $H=4$. We select $H=4$ as adding a further coordinate increases the area under the ROC curve by less than $0.01$, while $\mbox{AUC}_{4} -\mbox{AUC}_{3}>0.01$. In performing posterior inference we consider $5{,}000$ Gibbs samples with a burn-in of $1{,}000$. Convergence is assessed via visual inspection of selected traceplots and by \citet{gel_1992} potential scale reduction factors for the quantities of interest, obtaining comparable results to those in the simulation study. Using a four-dimensional latent space produces an area under the ROC curve for in-sample edge prediction of $0.978$. This is an interesting result in suggesting that the $120 \times 120$ time-varying adjacency matrices can be adequately characterized by collapsing information into a substantially lower-dimensional space. This insight is confirmed by results in Figures \ref{App0}--\ref{App1}, highlighting accurate performance in modeling dynamic network structures of interest.

\subsection{Posterior inference and model checking  in the application}
\begin{figure}
\centering
\includegraphics[trim=0.6cm 0.5cm 0.2cm 0.1cm, clip=true,height=7cm, width=15.5cm]{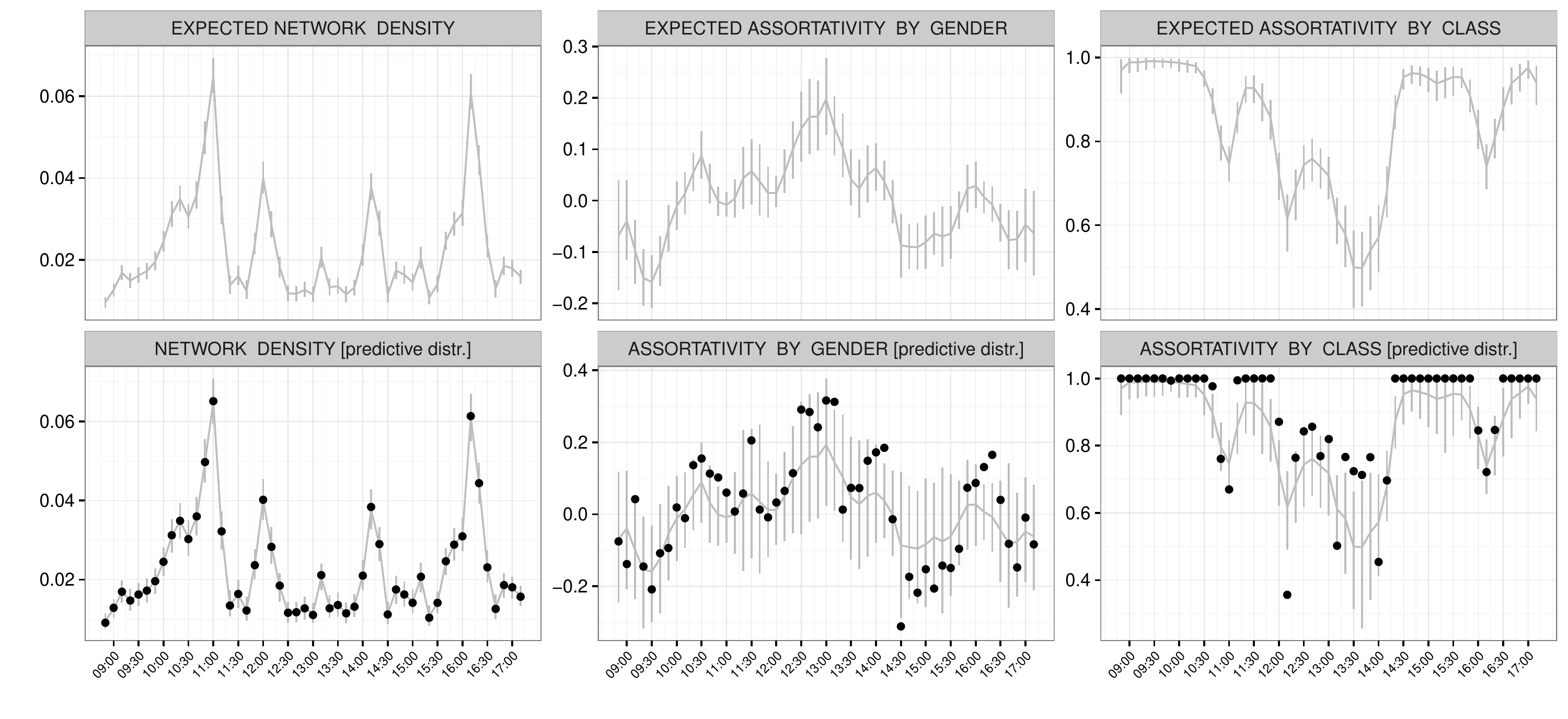}
\caption{\footnotesize{Upper panels: trajectory of the posterior mean (grey line) and point-wise $0.95$ highest posterior density intervals (grey segments) for dynamic expected network summary measures covering network density, assortativity by gender and assortativity by class. Bottom panels: for the same measures, mean trajectory (grey line) and point-wise $0.95$ predictive intervals (grey segments) obtained from the posterior predictive distribution; black dots represent the corresponding time-varying network measures computed from the observed data.}}
\label{App0}
\end{figure}

The trajectory of the posterior mean for the expected network density in the upper left plot of Figure \ref{App0} provides an interesting overview of the overall dynamic contact behavior, consistent with school schedule and changing environments summarized in Figure 10 of   \citet{steh_2011}. It is interesting to note how the expected network density evolves on low values suggesting a sparse network, with our adaptive procedure additionally capturing a rapid increase in the chance of contact occurring during school breaks and the beginning or the end of lunch times for groups of students. According to the left plot in the bottom panel of Figure \ref{App0}, the posterior predictive distribution arising from our formulation is sufficiently flexible in accommodating the evolution of these summary measures.

In studying dynamic homophily patterns, we investigate the posterior distribution of the time-varying  expected assortativity coefficients by class and gender, computed for the $115$ students. We hold out teachers in homophily studies as we don't have gender information for these actors and we are interested in social interactions among students --- consistent with  \citet{ste_2013}. In investigating gender homophily, \citet{ste_2013} focus on a single network obtained by aggregating the face-to-face interaction data that are observed in pre-selected nonconsecutive time windows when the occasions of contact are expected to have less environmental restrictions --- i.e. break and lunch times.  Although this is a reasonable procedure, information on spatial environments or events are not always available and the choice of aggregation intervals is not necessary unique. Moreover, investigating gender homophily for a single aggregated network provides only an averaged overview of a dynamic system. We instead study homophily structures as they evolve in time, and allow these quantities to be different in nonconsecutive time windows. 

Our results in the upper middle plot of Figure \ref{App0} partially confirm findings in \citet{ste_2013}, with the posterior distributions of the dynamic expected assortativity coefficients  concentrated on positive values during breaks and lunch times. However the expected assortativity is higher during lunch compared to breaks, with the posterior for these coefficients including the value $0$ during the last break. Hence  \citet{ste_2013} may over-estimate gender homophily in correspondence of breaks and under-estimate this property during lunches.

The expected assortativity by class is always positive, with the posterior distributions concentrating on substantially high values during school hours, when contacts are restricted by the spatial environments displayed in Figure 10 of   \citet{steh_2011}; refer to the upper right plot of Figure \ref{App0}. Model checking in the bottom middle and right plots of Figure \ref{App0} highlights an overall good performance of our procedures in characterizing also these higher-order homophily structures. These are key results, provided that we embed a $120 \times 120$ dynamic network into a substantially lower-dimensional space made by four latent coordinates, without any further information on the dynamic effect of exogenous variables. Few issues are found in accommodating rapid changes in assortativity by class. A reason behind this slight lack of fit is that  $H=4$ latent coordinates may not be sufficient to characterize class homophily in specific time windows. It is still an active area of research to accommodate latent space dimensions which adaptively change as a function of time. Similarly to our procedure, most available contributions rely on time-constant space dimensions. Although a subset of the observed class assortativity coefficients are not within the $0.95$ posterior predictive intervals, most of these values are contained in $0.99$ posterior predictive intervals. Hence, we maintain $H=4$ to avoid over-fitting. 

\begin{figure}
\centering
\includegraphics[trim=0.6cm 0.5cm 0.2cm 0.1cm, clip=true,height=7cm, width=15.5cm]{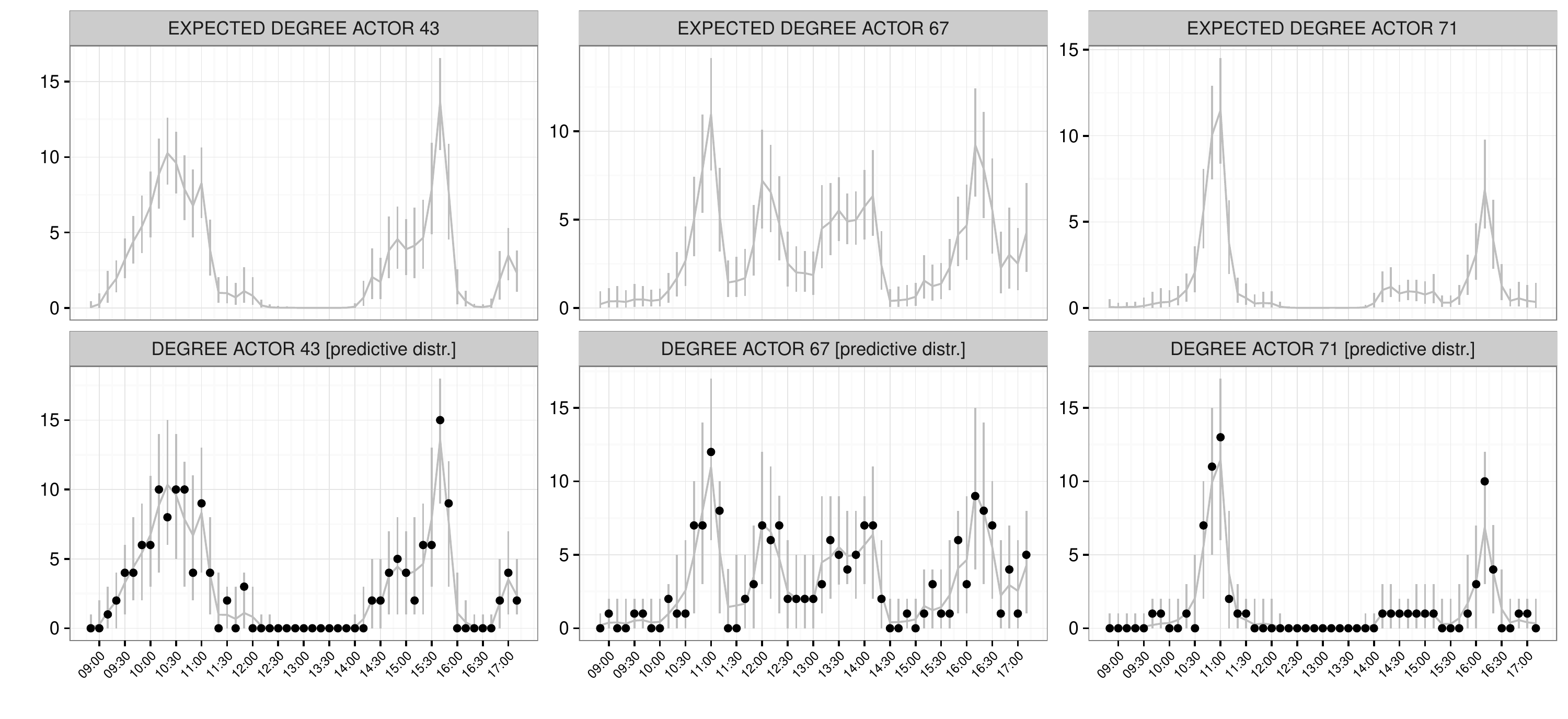}
\caption{\footnotesize{Upper panels: trajectory of the posterior mean (grey line) and point-wise $0.95$ highest posterior density intervals (grey segments) for the dynamic expected degree of selected actors. Bottom panels: for the same summary measures, mean trajectory (grey line) and point-wise $0.95$ predictive intervals (grey segments) obtained from the posterior predictive distribution; black dots represent the corresponding time-varying actors' degrees computed from the observed data.}}
\label{App1}
\end{figure}

Beside accommodating global network structures, our procedure can flexibly characterize actor-specific connectivity measures of interest. According to the upper panels of Figure \ref{App1}, incorporating across-actor heterogeneity and time-varying smoothness allows us to flexibly account for substantially different patterns and dynamic changes in expected actors' degrees. As shown in the bottom panels of Figure \ref{App1}, the posterior predictive distributions for the dynamic actors' degrees arising from our estimates are characterized by a very accurate performance in accommodating these time-varying observed quantities.

\begin{figure}
\centering
\includegraphics[trim=0.5cm 0.5cm 0cm 0cm, clip=true,height=17cm, width=15.5cm]{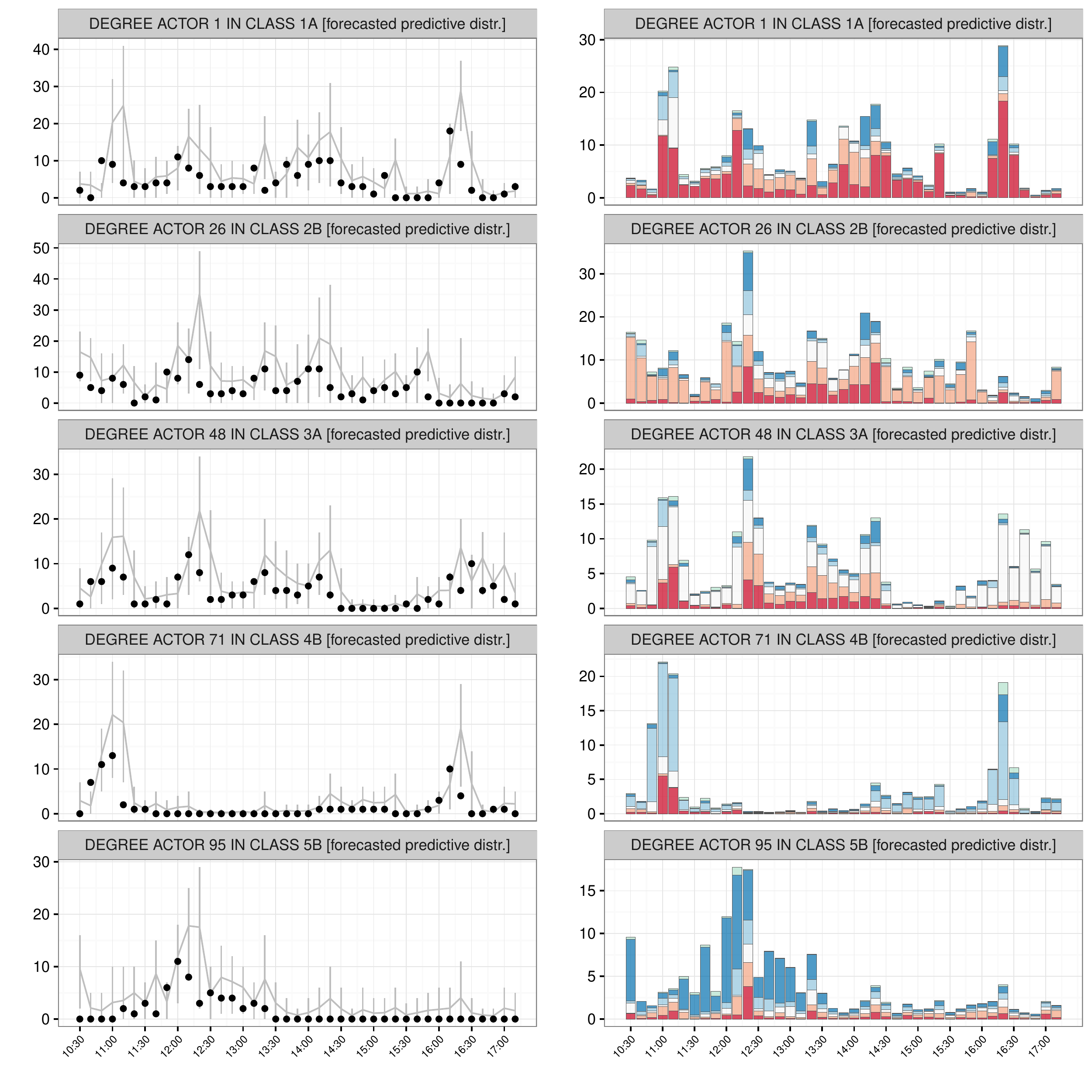}
\caption{\footnotesize{Left panels: for selected students in the five different classes,  mean trajectory (grey line) and point-wise $0.95$ predictive intervals (grey segments) of their degree obtained from the one-step-ahead forecasted predictive distribution from $t_{11}$ to $t_{51}$; black dots represents the corresponding time-varying actors' degrees computed from the observed data. Right panels: for the same students, barplots representing the  time-varying mean of their degree obtained from the one-step-ahead forecasted predictive distribution from $t_{11}$ to $t_{51}$. Colors in the bars represent the proportion of the forecasted degree due to connections with each class. Dark red (first class), light red (second class), white (third class), light blue (fourth class), dark blue (fifth class), green (teachers).}}
\label{App2}
\end{figure}

\subsection{Online updating, forecasting and predictive performance in the application}
\begin{figure}
\centering
\includegraphics[trim=0cm 0cm 0cm 0cm, clip=true,height=5cm, width=15.5cm]{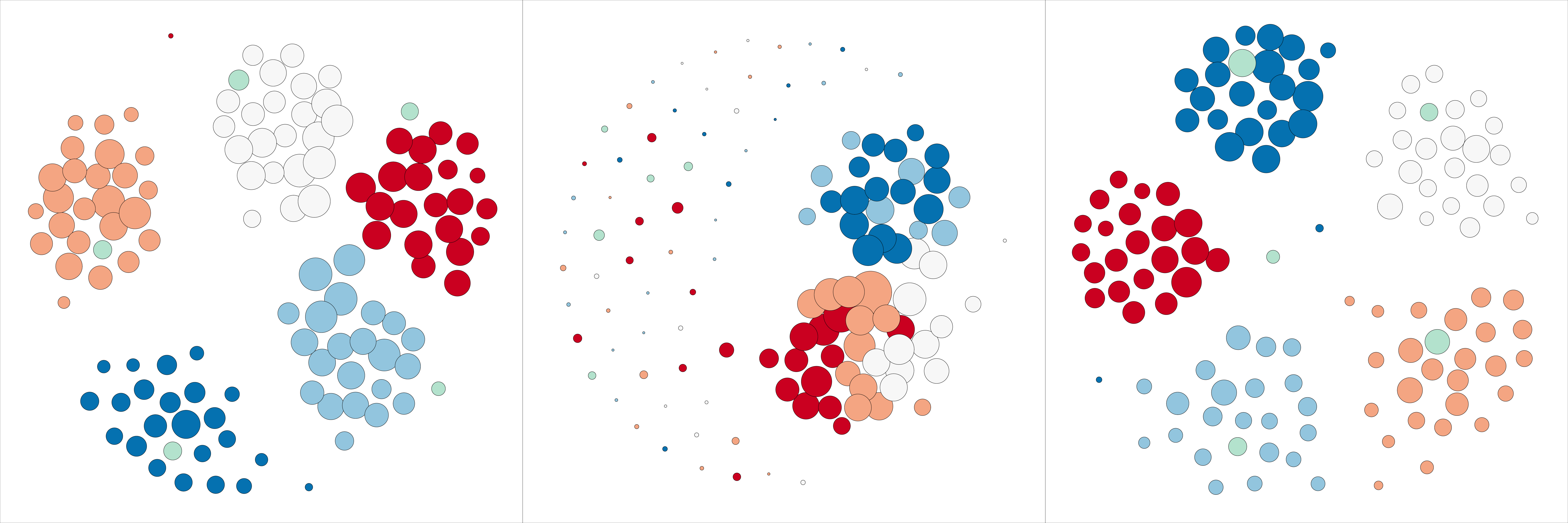}
\put (-440,145) {{\footnotesize{From $10{:}30$ to $12{:}00$}}}
\put (-293,145) {{\footnotesize{From $12{:}10$ to $14{:}00$}}}
\put (-145,145) {{\footnotesize{From $14{:}10$ to $17{:}10$}}}
\caption{\footnotesize{Weighted network visualization with weights obtained by averaging the mean of the one-step-ahead forecasted predictive distributions over three time windows. Edges are not displayed to facilitate graphical analysis. Actors' positions are obtained applying the \citet{fru_1991} force-directed placement algorithm, whereas their dimensions are proportional to the corresponding forecasted degree averaged over the three time windows.  The colors indicate class membership: dark red (first class), light red (second class), white (third class), light blue (fourth class), dark blue (fifth class), green (teachers).}}
\label{App3}
\end{figure}

\begin{figure}
\centering
\includegraphics[trim=0.5cm 0.7cm 0cm 0.2cm, clip=true,height=11cm, width=15cm]{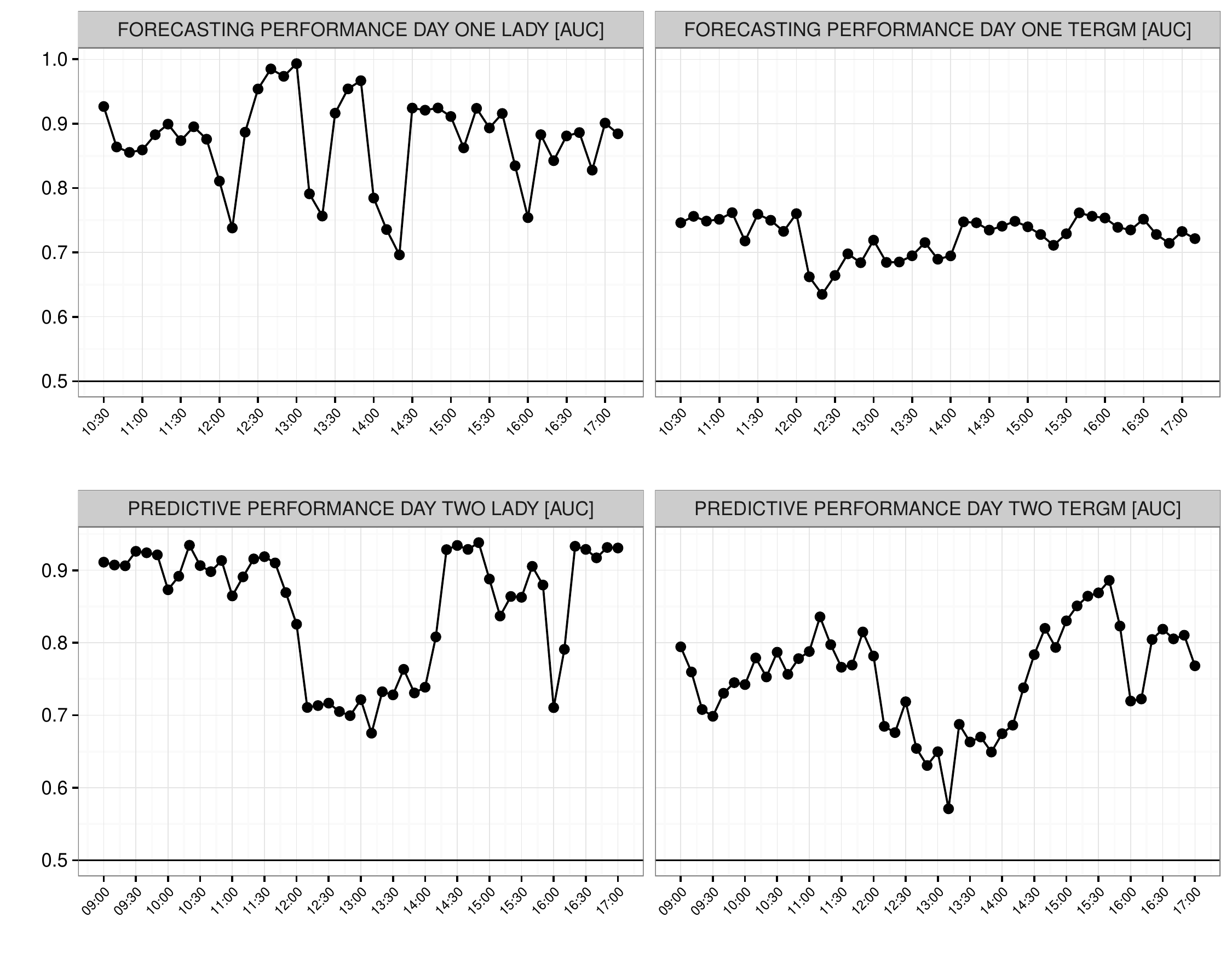}
\caption{\footnotesize{Upper panels: for times from $t_{11}$ to $t_{51}$ in day one, forecasting performance for our LADY network model and the TERGM. Performance is assessed via the areas under the ROC curves when forecasting the future edges with the mean of their corresponding one-step-ahead forecasted predictive distributions. Bottom panels: predictive performance for our LADY network model and the TERGM. Performance is assessed via the areas under the ROC curves when predicting edges in the networks $Y^*_{t_1}, \ldots, Y^*_{t_n}$  from day two, with the mean of their corresponding posterior predictive distribution estimated from the contact data in day one.}}
\label{App4}
\end{figure}

Once the model has been estimated on data from $t_1$ to $t_{i-1}$, a new contact network $Y_{t_{i}}$ can stream in along with the information that an individual --- or a subset of them --- has contracted a specific disease at $t_i$. For outbreak prevention, it is fundamental to rapidly update estimates at time $t_i$ and forecast the contact network structures at the next time $t_{i+1}$. Our LADY network model can suitably accomplish this task by online updating of the posterior distribution for the edge probabilities at $t_{i}$ exploiting strategies in Section \ref{online} --- with $j=5$ --- and then forecasting the posterior distribution of these probabilities at the next time $t_{i+1}$ by applying equation in Section \ref{fore} to the MCMC samples from the online updating. Once these quantities are available, it is easy to derive the approximate forecasted predictive distribution at $t_{i+1}$ along with related quantities of interest, such as its expected value for forecasting edges and the distribution  of future topological structures. Figures \ref{App2}, \ref{App3} and the upper left plot of Figure \ref{App4} evaluate the performance of our joint online updating and forecasting procedure for each $t_i=t_{10}, \ldots, t_{50}$, under different perspectives. 

The left panels of Figure \ref{App2} compare the observed degrees for selected students --- in the five different classes --- with their mean and quantiles arising from the forecasted predictive distribution. Time-varying actors' degrees are fundamental for disease surveillance and accurate forecasts for these quantities facilitates monitoring of the infectivity for each individual at future times. According to the left panels of Figure \ref{App2}, our strategies provide in general a good performance in forecasting dynamic degrees. We observe, however, a slight tendency towards over-estimating these quantities. Although this bias is undesirable, for the sake of outbreak prevention, slightly over-estimating actors' degrees  suggests conservative policies.

The right panels of  Figure \ref{App2} add further insights by showing the proportion of the forecasted degree due to connections with students in the different classes. This provides a higher-level measure of which groups of individuals are at risk of contagion at $t_{i+1}$ if a given individual contracts disease at $t_i$, for each $t_i=t_{10}, \ldots, t_{50}$. Results further confirm our good performance in forecasting heterogenous contact patterns and dynamic changes in actors degrees. Consistent with the findings on homophily, contacts with students from the same class represent a high proportion of the forecasted dynamic degrees. This is more evident during school hours than breaks or lunch times where we observe more mixed patterns including increased across class contacts and students apparently leaving the school --- such as for example actor $71$.

These findings are confirmed in Figure \ref{App3} providing a graphical representation of future networks with actors positions depending on the forecasted edges --- according to equation \eqref{for} --- averaged over three time windows of interest. Although we do not explicitly include environmental information, as shown in Figure \ref{App3} our procedure is sufficiently flexible to account for these structures from an unsupervised perspective. Consistent with Figure 10 in   \citet{steh_2011} we forecast evident community structures induced by class membership during the morning hours, with students in classes 1A, 3A and 4B being spatially closer than those in the remaining classes. This is consistent with classes 1A, 3A and 4B sharing the playground during the morning break according to Figure 10 in   \citet{steh_2011}. Lunch times are characterized by a sparse structure with two communities and a wide set of students having essentially no face-to-face contacts. The first community comprises students in classes 1A, 2B and part of those in class 3A. The second includes actors from classes 4B, 5B and the remaining students from class 3A. Also these forecasts are consistent with the approximate school schedule presented in \citet{steh_2011}, with a subset of the students leaving the school during lunch and the remaining individuals sharing the canteen in two different groups at consecutive times.  As expected, the results in the afternoon hours are similar to those in the morning, with a slightly more sparse structure due the fact that the students increasingly leave the school towards the end of the day.

The upper left plot in Figure \ref{App4} assesses forecasting performance by showing for each time from $t_{11}$ to $t_{51}$ the AUCs when forecasting the edges in each network  $Y_{t_{i+1}}$, for $t_{i+1}=t_{11}, \ldots, t_{51}$, with  the expectation of the forecasted predictive distribution --- estimated from data $Y_{t_{1}}, \ldots, Y_{t_{i}}$ under our  online updating and forecasting routine. The AUCs evolve on high values, suggesting an overall good performance in forecasting of future edges, with more evident decrements in correspondence of the beginning, mid and end of the lunch time windows. These times are characterized by rapid variations in contact behavior  due to students rapidly changing environments; refer to Figure 10 in \citet{steh_2011}.  Hence --- recalling also insights in the simulation study --- this decreased forecasting performance is reasonably related to the fact that the model has no experience of sudden regime changes. Although we face reduced forecasting performance in specific times, our procedure almost always improves forecasts from TERGM. Refer to the upper right plot in Figure \ref{App4}.

We conclude our analysis by evaluating the performance in predicting the edges of networks $Y^*_{t_1}, \ldots, Y^*_{t_n}$  during the second day,  based on the information provided by the face-to-face contact networks in the first day $Y_{t_1}, \ldots, Y_{t_n}$ --- consistent with discussion in Section \ref{fore}. In particular, we study the areas under the ROC curves when predicting the edges in each network $Y^*_{t_i}$ with the mean of their corresponding posterior predictive distribution  estimated from the contact data in day one; see equation \eqref{pred}. As time $t_{51}$ is not available in the second day, we assess out-of-sample predictive performance using data and estimates from $t_1$ to $t_{50}$. Results are displayed in the bottom panels of Figure  \ref{App4}.

We obtain a general good performance when predicting contacts in the second day, based on estimates from day one. More evident decrements  are found in correspondence of lunch times and the afternoon break. This may suggest that the dynamic contact networks at the second day are governed by slightly different underlying patterns than those associated with the first day, for these time windows. Also in this case we almost always improve results from the TERGM. These results further confirm the need of procedures accounting for heterogenous and dynamic dependence patterns in such frameworks.

\section{Discussion}
Although there has been an abundant interest in recent years in the study of dynamic networks, flexible methods for analyzing particular relational data have lagged behind the increasingly routine collection of such networks in several applied fields. Motivated by face-to-face dynamic contact networks, our methodology aims to take a further step towards addressing some of the current issues in dynamic network inference.

Our model has been constructed using latent similarity measures defined by the dot product of actor-specific latent coordinate vectors, with entries evolving in continuous time via nested Gaussian process to flexibly incorporate time-varying smoothness patterns. Using matrix factorization procedures, our LADY network model can accommodate moderately large $V$, and considering a state space representation of the nGP, we further allow scaling to larger time grids. Adapting  the P\'olya-gamma data augmentation strategy to our specific setting, we develop a simple and efficient Gibbs sampler for posterior computations, which utilizes standard results of Kalman filter for transformed Gaussian data. This further facilitates the development of forecasting, prediction and online updating procedures for fast inference.

Simulation studies confirm the good performance of the developed methodologies and the application allows us to learn interesting patterns in global and actor-specific network structures while confirming accurate forecasting and predictive performance. Recalling discussions in previous sections, there are several directions for future research. These include developing procedures to facilitate scaling to substantially larger sets of actors and improve model formulation to explicitly account for instantaneous and lagged covariates effects, without relying on potentially restrictive assumptions such as those typically encountered in dynamic ERGMs. Currently our model finds issues in scaling to very large networks and although our procedures have good inference and forecasting performance under an unsupervised perspective, careful inclusion of actors or edge covariates may further improve results.

We conclude our discussion by highlighting further fields of  application for our methodology. In fact, differently from TERGM and most of the available models specifically tailored for dynamic network inference, our LADY network model has a broader range of applicability also outside the temporal network field. In particular, our methods can be applied to network-valued data sets in which multiple observations of the same network are available along with a continuous predictor, instead of time. This is the case of neuroscience applications providing a network of structural interconnections among a common set of brain regions for different individuals along with  intelligence scores or personality traits --- among others. Replacing time with one of these predictors allows us to recast our LADY network model within a network regression framework which facilitates learning and prediction of changes in brain structural connectivity patterns across a cognitive trait of interest.

\section*{Acknowledgements}
We thank the Editor Edoardo M. Airoldi and the referees for the careful review process which helped us in substantially improving a first version of this paper. This work was partially funded by grant CPDA154381/15 of the University of Padova, Italy.

\bibliographystyle{imsart-nameyear}
\bibliography{daniele}

\end{document}